

\catcode`\@=11


\message{Loading jyTeX fonts...}



\font\vptrm=cmr5
\font\vptmit=cmmi5
\font\vptsy=cmsy5
\font\vptbf=cmbx5

\skewchar\vptmit='177 \skewchar\vptsy='60
\fontdimen16 \vptsy=\the\fontdimen17 \vptsy

\def\vpt{\ifmmode\err@badsizechange\else
     \@mathfontinit
     \textfont0=\vptrm  \scriptfont0=\vptrm  \scriptscriptfont0=\vptrm
     \textfont1=\vptmit \scriptfont1=\vptmit \scriptscriptfont1=\vptmit
     \textfont2=\vptsy  \scriptfont2=\vptsy  \scriptscriptfont2=\vptsy
     \textfont3=\xptex  \scriptfont3=\xptex  \scriptscriptfont3=\xptex
     \textfont\bffam=\vptbf
     \scriptfont\bffam=\vptbf
     \scriptscriptfont\bffam=\vptbf
     \@fontstyleinit
     \def\rm{\vptrm\fam=\z@}%
     \def\bf{\vptbf\fam=\bffam}%
     \def\oldstyle{\vptmit\fam=\@ne}%
     \rm\fi}


\font\viptrm=cmr6
\font\viptmit=cmmi6
\font\viptsy=cmsy6
\font\viptbf=cmbx6

\skewchar\viptmit='177 \skewchar\viptsy='60
\fontdimen16 \viptsy=\the\fontdimen17 \viptsy

\def\vipt{\ifmmode\err@badsizechange\else
     \@mathfontinit
     \textfont0=\viptrm  \scriptfont0=\vptrm  \scriptscriptfont0=\vptrm
     \textfont1=\viptmit \scriptfont1=\vptmit \scriptscriptfont1=\vptmit
     \textfont2=\viptsy  \scriptfont2=\vptsy  \scriptscriptfont2=\vptsy
     \textfont3=\xptex   \scriptfont3=\xptex  \scriptscriptfont3=\xptex
     \textfont\bffam=\viptbf
     \scriptfont\bffam=\vptbf
     \scriptscriptfont\bffam=\vptbf
     \@fontstyleinit
     \def\rm{\viptrm\fam=\z@}%
     \def\bf{\viptbf\fam=\bffam}%
     \def\oldstyle{\viptmit\fam=\@ne}%
     \rm\fi}


\font\viiptrm=cmr7
\font\viiptmit=cmmi7
\font\viiptsy=cmsy7
\font\viiptit=cmti7
\font\viiptbf=cmbx7

\skewchar\viiptmit='177 \skewchar\viiptsy='60
\fontdimen16 \viiptsy=\the\fontdimen17 \viiptsy

\def\viipt{\ifmmode\err@badsizechange\else
     \@mathfontinit
     \textfont0=\viiptrm  \scriptfont0=\vptrm  \scriptscriptfont0=\vptrm
     \textfont1=\viiptmit \scriptfont1=\vptmit \scriptscriptfont1=\vptmit
     \textfont2=\viiptsy  \scriptfont2=\vptsy  \scriptscriptfont2=\vptsy
     \textfont3=\xptex    \scriptfont3=\xptex  \scriptscriptfont3=\xptex
     \textfont\itfam=\viiptit
     \scriptfont\itfam=\viiptit
     \scriptscriptfont\itfam=\viiptit
     \textfont\bffam=\viiptbf
     \scriptfont\bffam=\vptbf
     \scriptscriptfont\bffam=\vptbf
     \@fontstyleinit
     \def\rm{\viiptrm\fam=\z@}%
     \def\it{\viiptit\fam=\itfam}%
     \def\bf{\viiptbf\fam=\bffam}%
     \def\oldstyle{\viiptmit\fam=\@ne}%
     \rm\fi}


\font\viiiptrm=cmr8
\font\viiiptmit=cmmi8
\font\viiiptsy=cmsy8
\font\viiiptit=cmti8
\font\viiiptbf=cmbx8

\skewchar\viiiptmit='177 \skewchar\viiiptsy='60
\fontdimen16 \viiiptsy=\the\fontdimen17 \viiiptsy

\def\viiipt{\ifmmode\err@badsizechange\else
     \@mathfontinit
     \textfont0=\viiiptrm  \scriptfont0=\viptrm  \scriptscriptfont0=\vptrm
     \textfont1=\viiiptmit \scriptfont1=\viptmit \scriptscriptfont1=\vptmit
     \textfont2=\viiiptsy  \scriptfont2=\viptsy  \scriptscriptfont2=\vptsy
     \textfont3=\xptex     \scriptfont3=\xptex   \scriptscriptfont3=\xptex
     \textfont\itfam=\viiiptit
     \scriptfont\itfam=\viiptit
     \scriptscriptfont\itfam=\viiptit
     \textfont\bffam=\viiiptbf
     \scriptfont\bffam=\viptbf
     \scriptscriptfont\bffam=\vptbf
     \@fontstyleinit
     \def\rm{\viiiptrm\fam=\z@}%
     \def\it{\viiiptit\fam=\itfam}%
     \def\bf{\viiiptbf\fam=\bffam}%
     \def\oldstyle{\viiiptmit\fam=\@ne}%
     \rm\fi}


\def\getixpt{%
     \font\ixptrm=cmr9
     \font\ixptmit=cmmi9
     \font\ixptsy=cmsy9
     \font\ixptit=cmti9
     \font\ixptbf=cmbx9
     \skewchar\ixptmit='177 \skewchar\ixptsy='60
     \fontdimen16 \ixptsy=\the\fontdimen17 \ixptsy}

\def\ixpt{\ifmmode\err@badsizechange\else
     \@mathfontinit
     \textfont0=\ixptrm  \scriptfont0=\viiptrm  \scriptscriptfont0=\vptrm
     \textfont1=\ixptmit \scriptfont1=\viiptmit \scriptscriptfont1=\vptmit
     \textfont2=\ixptsy  \scriptfont2=\viiptsy  \scriptscriptfont2=\vptsy
     \textfont3=\xptex   \scriptfont3=\xptex    \scriptscriptfont3=\xptex
     \textfont\itfam=\ixptit
     \scriptfont\itfam=\viiptit
     \scriptscriptfont\itfam=\viiptit
     \textfont\bffam=\ixptbf
     \scriptfont\bffam=\viiptbf
     \scriptscriptfont\bffam=\vptbf
     \@fontstyleinit
     \def\rm{\ixptrm\fam=\z@}%
     \def\it{\ixptit\fam=\itfam}%
     \def\bf{\ixptbf\fam=\bffam}%
     \def\oldstyle{\ixptmit\fam=\@ne}%
     \rm\fi}


\font\xptrm=cmr10
\font\xptmit=cmmi10
\font\xptsy=cmsy10
\font\xptex=cmex10
\font\xptit=cmti10
\font\xptsl=cmsl10
\font\xptbf=cmbx10
\font\xpttt=cmtt10
\font\xptss=cmss10
\font\xptsc=cmcsc10
\font\xptbfs=cmbx10
\font\xptbmit=cmmib10

\skewchar\xptmit='177 \skewchar\xptbmit='177 \skewchar\xptsy='60
\fontdimen16 \xptsy=\the\fontdimen17 \xptsy

\def\xpt{\ifmmode\err@badsizechange\else
     \@mathfontinit
     \textfont0=\xptrm  \scriptfont0=\viiptrm  \scriptscriptfont0=\vptrm
     \textfont1=\xptmit \scriptfont1=\viiptmit \scriptscriptfont1=\vptmit
     \textfont2=\xptsy  \scriptfont2=\viiptsy  \scriptscriptfont2=\vptsy
     \textfont3=\xptex  \scriptfont3=\xptex    \scriptscriptfont3=\xptex
     \textfont\itfam=\xptit
     \scriptfont\itfam=\viiptit
     \scriptscriptfont\itfam=\viiptit
     \textfont\bffam=\xptbf
     \scriptfont\bffam=\viiptbf
     \scriptscriptfont\bffam=\vptbf
     \textfont\bfsfam=\xptbfs
     \scriptfont\bfsfam=\viiptbf
     \scriptscriptfont\bfsfam=\vptbf
     \textfont\bmitfam=\xptbmit
     \scriptfont\bmitfam=\viiptmit
     \scriptscriptfont\bmitfam=\vptmit
     \@fontstyleinit
     \def\rm{\xptrm\fam=\z@}%
     \def\it{\xptit\fam=\itfam}%
     \def\sl{\xptsl}%
     \def\bf{\xptbf\fam=\bffam}%
     \def\tt{\xpttt}%
     \def\ss{\xptss}%
     \def\sc{\xptsc}%
     \def\bfs{\xptbfs\fam=\bfsfam}%
     \def\bmit{\fam=\bmitfam}%
     \def\oldstyle{\xptmit\fam=\@ne}%
     \rm\fi}


\def\getxipt{%
     \font\xiptrm=cmr10  scaled\magstephalf
     \font\xiptmit=cmmi10 scaled\magstephalf
     \font\xiptsy=cmsy10 scaled\magstephalf
     \font\xiptex=cmex10 scaled\magstephalf
     \font\xiptit=cmti10 scaled\magstephalf
     \font\xiptsl=cmsl10 scaled\magstephalf
     \font\xiptbf=cmbx10 scaled\magstephalf
     \font\xipttt=cmtt10 scaled\magstephalf
     \font\xiptss=cmss10 scaled\magstephalf
     \skewchar\xiptmit='177 \skewchar\xiptsy='60
     \fontdimen16 \xiptsy=\the\fontdimen17 \xiptsy}

\def\xipt{\ifmmode\err@badsizechange\else
     \@mathfontinit
     \textfont0=\xiptrm  \scriptfont0=\viiiptrm  \scriptscriptfont0=\viptrm
     \textfont1=\xiptmit \scriptfont1=\viiiptmit \scriptscriptfont1=\viptmit
     \textfont2=\xiptsy  \scriptfont2=\viiiptsy  \scriptscriptfont2=\viptsy
     \textfont3=\xiptex  \scriptfont3=\xptex     \scriptscriptfont3=\xptex
     \textfont\itfam=\xiptit
     \scriptfont\itfam=\viiiptit
     \scriptscriptfont\itfam=\viiptit
     \textfont\bffam=\xiptbf
     \scriptfont\bffam=\viiiptbf
     \scriptscriptfont\bffam=\viptbf
     \@fontstyleinit
     \def\rm{\xiptrm\fam=\z@}%
     \def\it{\xiptit\fam=\itfam}%
     \def\sl{\xiptsl}%
     \def\bf{\xiptbf\fam=\bffam}%
     \def\tt{\xipttt}%
     \def\ss{\xiptss}%
     \def\oldstyle{\xiptmit\fam=\@ne}%
     \rm\fi}


\font\xiiptrm=cmr12
\font\xiiptmit=cmmi12
\font\xiiptsy=cmsy10  scaled\magstep1
\font\xiiptex=cmex10  scaled\magstep1
\font\xiiptit=cmti12
\font\xiiptsl=cmsl12
\font\xiiptbf=cmbx12
\font\xiipttt=cmtt12
\font\xiiptss=cmss12
\font\xiiptsc=cmcsc10 scaled\magstep1
\font\xiiptbfs=cmbx10  scaled\magstep1
\font\xiiptbmit=cmmib10 scaled\magstep1

\skewchar\xiiptmit='177 \skewchar\xiiptbmit='177 \skewchar\xiiptsy='60
\fontdimen16 \xiiptsy=\the\fontdimen17 \xiiptsy

\def\xiipt{\ifmmode\err@badsizechange\else
     \@mathfontinit
     \textfont0=\xiiptrm  \scriptfont0=\viiiptrm  \scriptscriptfont0=\viptrm
     \textfont1=\xiiptmit \scriptfont1=\viiiptmit \scriptscriptfont1=\viptmit
     \textfont2=\xiiptsy  \scriptfont2=\viiiptsy  \scriptscriptfont2=\viptsy
     \textfont3=\xiiptex  \scriptfont3=\xptex     \scriptscriptfont3=\xptex
     \textfont\itfam=\xiiptit
     \scriptfont\itfam=\viiiptit
     \scriptscriptfont\itfam=\viiptit
     \textfont\bffam=\xiiptbf
     \scriptfont\bffam=\viiiptbf
     \scriptscriptfont\bffam=\viptbf
     \textfont\bfsfam=\xiiptbfs
     \scriptfont\bfsfam=\viiiptbf
     \scriptscriptfont\bfsfam=\viptbf
     \textfont\bmitfam=\xiiptbmit
     \scriptfont\bmitfam=\viiiptmit
     \scriptscriptfont\bmitfam=\viptmit
     \@fontstyleinit
     \def\rm{\xiiptrm\fam=\z@}%
     \def\it{\xiiptit\fam=\itfam}%
     \def\sl{\xiiptsl}%
     \def\bf{\xiiptbf\fam=\bffam}%
     \def\tt{\xiipttt}%
     \def\ss{\xiiptss}%
     \def\sc{\xiiptsc}%
     \def\bfs{\xiiptbfs\fam=\bfsfam}%
     \def\bmit{\fam=\bmitfam}%
     \def\oldstyle{\xiiptmit\fam=\@ne}%
     \rm\fi}


\def\getxiiipt{%
     \font\xiiiptrm=cmr12  scaled\magstephalf
     \font\xiiiptmit=cmmi12 scaled\magstephalf
     \font\xiiiptsy=cmsy9  scaled\magstep2
     \font\xiiiptit=cmti12 scaled\magstephalf
     \font\xiiiptsl=cmsl12 scaled\magstephalf
     \font\xiiiptbf=cmbx12 scaled\magstephalf
     \font\xiiipttt=cmtt12 scaled\magstephalf
     \font\xiiiptss=cmss12 scaled\magstephalf
     \skewchar\xiiiptmit='177 \skewchar\xiiiptsy='60
     \fontdimen16 \xiiiptsy=\the\fontdimen17 \xiiiptsy}

\def\xiiipt{\ifmmode\err@badsizechange\else
     \@mathfontinit
     \textfont0=\xiiiptrm  \scriptfont0=\xptrm  \scriptscriptfont0=\viiptrm
     \textfont1=\xiiiptmit \scriptfont1=\xptmit \scriptscriptfont1=\viiptmit
     \textfont2=\xiiiptsy  \scriptfont2=\xptsy  \scriptscriptfont2=\viiptsy
     \textfont3=\xivptex   \scriptfont3=\xptex  \scriptscriptfont3=\xptex
     \textfont\itfam=\xiiiptit
     \scriptfont\itfam=\xptit
     \scriptscriptfont\itfam=\viiptit
     \textfont\bffam=\xiiiptbf
     \scriptfont\bffam=\xptbf
     \scriptscriptfont\bffam=\viiptbf
     \@fontstyleinit
     \def\rm{\xiiiptrm\fam=\z@}%
     \def\it{\xiiiptit\fam=\itfam}%
     \def\sl{\xiiiptsl}%
     \def\bf{\xiiiptbf\fam=\bffam}%
     \def\tt{\xiiipttt}%
     \def\ss{\xiiiptss}%
     \def\oldstyle{\xiiiptmit\fam=\@ne}%
     \rm\fi}


\font\xivptrm=cmr12   scaled\magstep1
\font\xivptmit=cmmi12  scaled\magstep1
\font\xivptsy=cmsy10  scaled\magstep2
\font\xivptex=cmex10  scaled\magstep2
\font\xivptit=cmti12  scaled\magstep1
\font\xivptsl=cmsl12  scaled\magstep1
\font\xivptbf=cmbx12  scaled\magstep1
\font\xivpttt=cmtt12  scaled\magstep1
\font\xivptss=cmss12  scaled\magstep1
\font\xivptsc=cmcsc10 scaled\magstep2
\font\xivptbfs=cmbx10  scaled\magstep2
\font\xivptbmit=cmmib10 scaled\magstep2

\skewchar\xivptmit='177 \skewchar\xivptbmit='177 \skewchar\xivptsy='60
\fontdimen16 \xivptsy=\the\fontdimen17 \xivptsy

\def\xivpt{\ifmmode\err@badsizechange\else
     \@mathfontinit
     \textfont0=\xivptrm  \scriptfont0=\xptrm  \scriptscriptfont0=\viiptrm
     \textfont1=\xivptmit \scriptfont1=\xptmit \scriptscriptfont1=\viiptmit
     \textfont2=\xivptsy  \scriptfont2=\xptsy  \scriptscriptfont2=\viiptsy
     \textfont3=\xivptex  \scriptfont3=\xptex  \scriptscriptfont3=\xptex
     \textfont\itfam=\xivptit
     \scriptfont\itfam=\xptit
     \scriptscriptfont\itfam=\viiptit
     \textfont\bffam=\xivptbf
     \scriptfont\bffam=\xptbf
     \scriptscriptfont\bffam=\viiptbf
     \textfont\bfsfam=\xivptbfs
     \scriptfont\bfsfam=\xptbfs
     \scriptscriptfont\bfsfam=\viiptbf
     \textfont\bmitfam=\xivptbmit
     \scriptfont\bmitfam=\xptbmit
     \scriptscriptfont\bmitfam=\viiptmit
     \@fontstyleinit
     \def\rm{\xivptrm\fam=\z@}%
     \def\it{\xivptit\fam=\itfam}%
     \def\sl{\xivptsl}%
     \def\bf{\xivptbf\fam=\bffam}%
     \def\tt{\xivpttt}%
     \def\ss{\xivptss}%
     \def\sc{\xivptsc}%
     \def\bfs{\xivptbfs\fam=\bfsfam}%
     \def\bmit{\fam=\bmitfam}%
     \def\oldstyle{\xivptmit\fam=\@ne}%
     \rm\fi}


\font\xviiptrm=cmr17
\font\xviiptmit=cmmi12 scaled\magstep2
\font\xviiptsy=cmsy10 scaled\magstep3
\font\xviiptex=cmex10 scaled\magstep3
\font\xviiptit=cmti12 scaled\magstep2
\font\xviiptbf=cmbx12 scaled\magstep2
\font\xviiptbfs=cmbx10 scaled\magstep3

\skewchar\xviiptmit='177 \skewchar\xviiptsy='60
\fontdimen16 \xviiptsy=\the\fontdimen17 \xviiptsy

\def\xviipt{\ifmmode\err@badsizechange\else
     \@mathfontinit
     \textfont0=\xviiptrm  \scriptfont0=\xiiptrm  \scriptscriptfont0=\viiiptrm
     \textfont1=\xviiptmit \scriptfont1=\xiiptmit \scriptscriptfont1=\viiiptmit
     \textfont2=\xviiptsy  \scriptfont2=\xiiptsy  \scriptscriptfont2=\viiiptsy
     \textfont3=\xviiptex  \scriptfont3=\xiiptex  \scriptscriptfont3=\xptex
     \textfont\itfam=\xviiptit
     \scriptfont\itfam=\xiiptit
     \scriptscriptfont\itfam=\viiiptit
     \textfont\bffam=\xviiptbf
     \scriptfont\bffam=\xiiptbf
     \scriptscriptfont\bffam=\viiiptbf
     \textfont\bfsfam=\xviiptbfs
     \scriptfont\bfsfam=\xiiptbfs
     \scriptscriptfont\bfsfam=\viiiptbf
     \@fontstyleinit
     \def\rm{\xviiptrm\fam=\z@}%
     \def\it{\xviiptit\fam=\itfam}%
     \def\bf{\xviiptbf\fam=\bffam}%
     \def\bfs{\xviiptbfs\fam=\bfsfam}%
     \def\oldstyle{\xviiptmit\fam=\@ne}%
     \rm\fi}


\font\xxiptrm=cmr17  scaled\magstep1


\def\xxipt{\ifmmode\err@badsizechange\else
     \@mathfontinit
     \@fontstyleinit
     \def\rm{\xxiptrm\fam=\z@}%
     \rm\fi}


\font\xxvptrm=cmr17  scaled\magstep2


\def\xxvpt{\ifmmode\err@badsizechange\else
     \@mathfontinit
     \@fontstyleinit
     \def\rm{\xxvptrm\fam=\z@}%
     \rm\fi}




\message{Loading jyTeX macros...}

\message{modifications to plain.tex,}


\def\newcount{\alloc@0\count\countdef\insc@unt}
\def\newdimen{\alloc@1\dimen\dimendef\insc@unt}
\def\newskip{\alloc@2\skip\skipdef\insc@unt}
\def\newmuskip{\alloc@3\muskip\muskipdef\@cclvi}
\def\newbox{\alloc@4\box\chardef\insc@unt}
\def\newtoks{\alloc@5\toks\toksdef\@cclvi}
\def\newhelp#1#2{\newtoks#1\global#1\expandafter{\csname#2\endcsname}}
\def\newread{\alloc@6\read\chardef\sixt@@n}
\def\newwrite{\alloc@7\write\chardef\sixt@@n}
\def\newfam{\alloc@8\fam\chardef\sixt@@n}
\def\newinsert#1{\global\advance\insc@unt by\m@ne
     \ch@ck0\insc@unt\count
     \ch@ck1\insc@unt\dimen
     \ch@ck2\insc@unt\skip
     \ch@ck4\insc@unt\box
     \allocationnumber=\insc@unt
     \global\chardef#1=\allocationnumber
     \wlog{\string#1=\string\insert\the\allocationnumber}}
\def\newif#1{\count@\escapechar \escapechar\m@ne
     \expandafter\expandafter\expandafter
          \xdef\@if#1{true}{\let\noexpand#1=\noexpand\iftrue}%
     \expandafter\expandafter\expandafter
          \xdef\@if#1{false}{\let\noexpand#1=\noexpand\iffalse}%
     \global\@if#1{false}\escapechar=\count@}


\newlinechar=`\^^J
\overfullrule=0pt




\let\itfam=\undefined

\let\bffam=\undefined

\count18=3


\chardef\sharps="19


\mathchardef\alpha="710B
\mathchardef\beta="710C
\mathchardef\gamma="710D
\mathchardef\delta="710E
\mathchardef\epsilon="710F
\mathchardef\zeta="7110
\mathchardef\eta="7111
\mathchardef\theta="7112
\mathchardef\iota="7113
\mathchardef\kappa="7114
\mathchardef\lambda="7115
\mathchardef\mu="7116
\mathchardef\nu="7117
\mathchardef\xi="7118
\mathchardef\pi="7119
\mathchardef\rho="711A
\mathchardef\sigma="711B
\mathchardef\tau="711C
\mathchardef\upsilon="711D
\mathchardef\phi="711E
\mathchardef\chi="711F
\mathchardef\psi="7120
\mathchardef\omega="7121
\mathchardef\varepsilon="7122
\mathchardef\vartheta="7123
\mathchardef\varpi="7124
\mathchardef\varrho="7125
\mathchardef\varsigma="7126
\mathchardef\varphi="7127
\mathchardef\imath="717B
\mathchardef\jmath="717C
\mathchardef\ell="7160
\mathchardef\wp="717D
\mathchardef\partial="7140
\mathchardef\flat="715B
\mathchardef\natural="715C
\mathchardef\sharp="715D



\def\angle{{\vbox{\ialign{$\m@th\scriptstyle##$\crcr
     \not\mathrel{\mkern14mu}\crcr
     \noalign{\nointerlineskip}
     \mkern2.5mu\leaders\hrule height.34\rp@\hfill\mkern2.5mu\crcr}}}}
\def\vdots{\vbox{\baselineskip4\rp@ \lineskiplimit\z@
     \kern6\rp@\hbox{.}\hbox{.}\hbox{.}}}
\def\ddots{\mathinner{\mkern1mu\raise7\rp@\vbox{\kern7\rp@\hbox{.}}\mkern2mu
     \raise4\rp@\hbox{.}\mkern2mu\raise\rp@\hbox{.}\mkern1mu}}
\def\overrightarrow#1{\vbox{\ialign{##\crcr
     \rightarrowfill\crcr
     \noalign{\kern-\rp@\nointerlineskip}
     $\hfil\displaystyle{#1}\hfil$\crcr}}}
\def\overleftarrow#1{\vbox{\ialign{##\crcr
     \leftarrowfill\crcr
     \noalign{\kern-\rp@\nointerlineskip}
     $\hfil\displaystyle{#1}\hfil$\crcr}}}
\def\overbrace#1{\mathop{\vbox{\ialign{##\crcr
     \noalign{\kern3\rp@}
     \downbracefill\crcr
     \noalign{\kern3\rp@\nointerlineskip}
     $\hfil\displaystyle{#1}\hfil$\crcr}}}\limits}
\def\underbrace#1{\mathop{\vtop{\ialign{##\crcr
     $\hfil\displaystyle{#1}\hfil$\crcr
     \noalign{\kern3\rp@\nointerlineskip}
     \upbracefill\crcr
     \noalign{\kern3\rp@}}}}\limits}
\def\big#1{{\hbox{$\left#1\vbox to8.5\rp@ {}\right.\n@space$}}}
\def\Big#1{{\hbox{$\left#1\vbox to11.5\rp@ {}\right.\n@space$}}}
\def\bigg#1{{\hbox{$\left#1\vbox to14.5\rp@ {}\right.\n@space$}}}
\def\Bigg#1{{\hbox{$\left#1\vbox to17.5\rp@ {}\right.\n@space$}}}
\def\@vereq#1#2{\lower.5\rp@\vbox{\baselineskip\z@skip\lineskip-.5\rp@
     \ialign{$\m@th#1\hfil##\hfil$\crcr#2\crcr=\crcr}}}
\def\rlh@#1{\vcenter{\hbox{\ooalign{\raise2\rp@
     \hbox{$#1\rightharpoonup$}\crcr
     $#1\leftharpoondown$}}}}
\def\bordermatrix#1{\begingroup\m@th
     \setbox\z@\vbox{%
          \def\cr{\crcr\noalign{\kern2\rp@\global\let\cr\endline}}%
          \ialign{$##$\hfil\kern2\rp@\kern\p@renwd
               &\thinspace\hfil$##$\hfil&&\quad\hfil$##$\hfil\crcr
               \omit\strut\hfil\crcr
               \noalign{\kern-\baselineskip}%
               #1\crcr\omit\strut\cr}}%
     \setbox\tw@\vbox{\unvcopy\z@\global\setbox\@ne\lastbox}%
     \setbox\tw@\hbox{\unhbox\@ne\unskip\global\setbox\@ne\lastbox}%
     \setbox\tw@\hbox{$\kern\wd\@ne\kern-\p@renwd\left(\kern-\wd\@ne
          \global\setbox\@ne\vbox{\box\@ne\kern2\rp@}%
          \vcenter{\kern-\ht\@ne\unvbox\z@\kern-\baselineskip}%
          \,\right)$}%
     \null\;\vbox{\kern\ht\@ne\box\tw@}\endgroup}
\def\endinsert{\egroup
     \if@mid\dimen@\ht\z@
          \advance\dimen@\dp\z@
          \advance\dimen@12\rp@
          \advance\dimen@\pagetotal
          \ifdim\dimen@>\pagegoal\@midfalse\p@gefalse\fi
     \fi
     \if@mid\bigskip\box\z@
          \bigbreak
     \else\insert\topins{\penalty100 \splittopskip\z@skip
               \splitmaxdepth\maxdimen\floatingpenalty\z@
               \ifp@ge\dimen@\dp\z@
                    \vbox to\vsize{\unvbox\z@\kern-\dimen@}%
               \else\box\z@\nobreak\bigskip
               \fi}%
     \fi
     \endgroup}


\def\cases#1{\left\{\,\vcenter{\m@th
     \ialign{$##\hfil$&\quad##\hfil\crcr#1\crcr}}\right.}
\def\matrix#1{\null\,\vcenter{\m@th
     \ialign{\hfil$##$\hfil&&\quad\hfil$##$\hfil\crcr
          \mathstrut\crcr
          \noalign{\kern-\baselineskip}
          #1\crcr
          \mathstrut\crcr
          \noalign{\kern-\baselineskip}}}\,}


\newif\ifraggedbottom

\def\raggedbottom{\ifraggedbottom\else
     \advance\topskip by\z@ plus60pt \raggedbottomtrue\fi}%
\def\normalbottom{\ifraggedbottom
     \advance\topskip by\z@ plus-60pt \raggedbottomfalse\fi}

\message{hacks,}


\toksdef\toks@i=1
\toksdef\toks@ii=2


\def\TeX{T\kern-.1667em \lower.5ex \hbox{E}\kern-.125em X\null}
\def\jyTeX{{\leavevmode
     \raise.587ex \hbox{\it\j}\kern-.1em \lower.048ex \hbox{\it y}\kern-.12em
     \TeX}}

\let\then=\iftrue
\def\ifnoarg#1\then{\def\hack@{#1}\ifx\hack@\empty}
\def\ifundefined#1\then{%
     \expandafter\ifx\csname\expandafter\blank\string#1\endcsname\relax}
\def\useif#1\then{\csname#1\endcsname}
\def\usename#1{\csname#1\endcsname}
\def\useafter#1#2{\expandafter#1\csname#2\endcsname}

\long\def\loop#1\repeat{\def\@iterate{#1\expandafter\@iterate\fi}\@iterate
     \let\@iterate=\relax}

\let\TeXend=\end
\def\begin#1{\begingroup\def\@@blockname{#1}\usename{begin#1}}
\def\end#1{\usename{end#1}\def\hack@{#1}%
     \ifx\@@blockname\hack@
          \endgroup
     \else\err@badgroup\hack@\@@blockname
     \fi}
\def\@@blockname{}

\def\defaultoption[#1]#2{%
     \def\hack@{\ifx\hack@ii[\toks@={#2}\else\toks@={#2[#1]}\fi\the\toks@}%
     \futurelet\hack@ii\hack@}

\def\markup#1{\let\@@marksf=\empty
     \ifhmode\edef\@@marksf{\spacefactor=\the\spacefactor\relax}\/\fi
     ${}^{\hbox{\subscriptfonts#1}}$\@@marksf}


\newtoks\shortyear
\newtoks\militaryhour
\newtoks\standardhour
\newtoks\minute
\newtoks\amorpm

\def\settime{\count@=\time\divide\count@ by60
     \militaryhour=\expandafter{\number\count@}%
     {\multiply\count@ by-60 \advance\count@ by\time
          \xdef\hack@{\ifnum\count@<10 0\fi\number\count@}}%
     \minute=\expandafter{\hack@}%
     \ifnum\count@<12
          \amorpm={am}
     \else\amorpm={pm}
          \ifnum\count@>12 \advance\count@ by-12 \fi
     \fi
     \standardhour=\expandafter{\number\count@}%
     \def\hack@19##1##2{\shortyear={##1##2}}%
          \expandafter\hack@\the\year}

\def\monthword#1{%
     \ifcase#1
          $\bullet$\err@badcountervalue{monthword}%
          \or January\or February\or March\or April\or May\or June%
          \or July\or August\or September\or October\or November\or December%
     \else$\bullet$\err@badcountervalue{monthword}%
     \fi}

\def\monthabbr#1{%
     \ifcase#1
          $\bullet$\err@badcountervalue{monthabbr}%
          \or Jan\or Feb\or Mar\or Apr\or May\or Jun%
          \or Jul\or Aug\or Sep\or Oct\or Nov\or Dec%
     \else$\bullet$\err@badcountervalue{monthabbr}%
     \fi}

\def\militarytime{\the\militaryhour:\the\minute}
\def\standardtime{\the\standardhour:\the\minute}


\def\@setnumstyle#1#2{\expandafter\global\expandafter\expandafter
     \expandafter\let\expandafter\expandafter
     \csname @\expandafter\blank\string#1style\endcsname
     \csname#2\endcsname}
\def\numstyle#1{\usename{@\expandafter\blank\string#1style}#1}
\def\ifblank#1\then{\useafter\ifx{@\expandafter\blank\string#1}\blank}

\def\blank#1{}

\def\Roman#1{\expandafter\uppercase\expandafter{\romannumeral#1}}
\def\alphabetic#1{%
     \ifcase#1
          $\bullet$\err@badcountervalue{alphabetic}%
          \or a\or b\or c\or d\or e\or f\or g\or h\or i\or j\or k\or l\or m%
          \or n\or o\or p\or q\or r\or s\or t\or u\or v\or w\or x\or y\or z%
     \else$\bullet$\err@badcountervalue{alphabetic}%
     \fi}
\def\Alphabetic#1{\expandafter\uppercase\expandafter{\alphabetic{#1}}}
\def\symbols#1{%
     \ifcase#1
          $\bullet$\err@badcountervalue{symbols}%
          \or*\or\dag\or\ddag\or\S\or$\|$%
          \or**\or\dag\dag\or\ddag\ddag\or\S\S\or$\|\|$%
     \else$\bullet$\err@badcountervalue{symbols}%
     \fi}


\catcode`\^^?=13 \def^^?{\relax}

\def\trimleading#1\to#2{\edef#2{#1}%
     \expandafter\@trimleading\expandafter#2#2^^?^^?}
\def\@trimleading#1#2#3^^?{\ifx#2^^?\def#1{}\else\def#1{#2#3}\fi}

\def\trimtrailing#1\to#2{\edef#2{#1}%
     \expandafter\@trimtrailing\expandafter#2#2^^? ^^?\relax}
\def\@trimtrailing#1#2 ^^?#3{\ifx#3\relax\toks@={}%
     \else\def#1{#2}\toks@={\trimtrailing#1\to#1}\fi
     \the\toks@}

\def\trim#1\to#2{\trimleading#1\to#2\trimtrailing#2\to#2}

\catcode`\^^?=15


\long\def\additemL#1\to#2{\toks@={\^^\{#1}}\toks@ii=\expandafter{#2}%
     \xdef#2{\the\toks@\the\toks@ii}}

\long\def\additemR#1\to#2{\toks@={\^^\{#1}}\toks@ii=\expandafter{#2}%
     \xdef#2{\the\toks@ii\the\toks@}}

\def\getitemL#1\to#2{\expandafter\@getitemL#1\hack@#1#2}
\def\@getitemL\^^\#1#2\hack@#3#4{\def#4{#1}\def#3{#2}}

\message{font macros,}


\newdimen\rp@
\newcount\@@sizeindex \@@sizeindex=0
\newcount\@@factori
\newcount\@@factorii
\newcount\@@factoriii
\newcount\@@factoriv

\countdef\maxfam=18
\newfam\itfam
\newfam\bffam
\newfam\bfsfam
\newfam\bmitfam

\def\@mathfontinit{\count@=4
     \loop\textfont\count@=\nullfont
          \scriptfont\count@=\nullfont
          \scriptscriptfont\count@=\nullfont
          \ifnum\count@<\maxfam\advance\count@ by\@ne
     \repeat}

\def\@fontstyleinit{%
     \def\it{\err@fontnotavailable\it}%
     \def\bf{\err@fontnotavailable\bf}%
     \def\bfs{\err@bfstobf}%
     \def\bmit{\err@fontnotavailable\bmit}%
     \def\sc{\err@fontnotavailable\sc}%
     \def\sl{\err@sltoit}%
     \def\ss{\err@fontnotavailable\ss}%
     \def\tt{\err@fontnotavailable\tt}}

\def\@parameterinit#1{\rm\rp@=.1em \@getscaling{#1}%
     \let\^^\=\@doscaling\scalingskipslist
     \setbox\strutbox=\hbox{\vrule
          height.708\baselineskip depth.292\baselineskip width\z@}}

\def\@getfactor#1#2#3#4{\@@factori=#1 \@@factorii=#2
     \@@factoriii=#3 \@@factoriv=#4}

\def\@getscaling#1{\count@=#1 \advance\count@ by-\@@sizeindex\@@sizeindex=#1
     \ifnum\count@<0
          \let\@mulordiv=\divide
          \let\@divormul=\multiply
          \multiply\count@ by\m@ne
     \else\let\@mulordiv=\multiply
          \let\@divormul=\divide
     \fi
     \edef\@@scratcha{\ifcase\count@                {1}{1}{1}{1}\or
          {1}{7}{23}{3}\or     {2}{5}{3}{1}\or      {9}{89}{13}{1}\or
          {6}{25}{6}{1}\or     {8}{71}{14}{1}\or    {6}{25}{36}{5}\or
          {1}{7}{53}{4}\or     {12}{125}{108}{5}\or {3}{14}{53}{5}\or
          {6}{41}{17}{1}\or    {13}{31}{13}{2}\or   {9}{107}{71}{2}\or
          {11}{139}{124}{3}\or {1}{6}{43}{2}\or     {10}{107}{42}{1}\or
          {1}{5}{43}{2}\or     {5}{69}{65}{1}\or    {11}{97}{91}{2}\fi}%
     \expandafter\@getfactor\@@scratcha}

\def\@doscaling#1{\@mulordiv#1by\@@factori\@divormul#1by\@@factorii
     \@mulordiv#1by\@@factoriii\@divormul#1by\@@factoriv}


\newskip\headskip
\newskip\footskip

\def\typesize=#1pt{\count@=#1 \advance\count@ by-10
     \ifcase\count@
          \@setsizex\or\err@badtypesize\or
          \@setsizexii\or\err@badtypesize\or
          \@setsizexiv
     \else\err@badtypesize
     \fi}

\def\@setsizex{\getixpt
     \def\subsubscriptfonts{\vpt}%
          \def\subsubscriptsize{\vpt\@parameterinit{-8}}%
     \def\subscriptfonts{\viipt}\def\subscriptsize{\viipt\@parameterinit{-4}}%
     \def\footnotefonts{\viiipt}\def\footnotesize{\viiipt\@parameterinit{-2}}%
     \def\smallfonts{\ixpt}\def\smallsize{\ixpt\@parameterinit{-1}}%
     \def\normalfonts{\xpt}\def\normalsize{\xpt\@parameterinit{0}}%
     \def\bigfonts{\xiipt}\def\bigsize{\xiipt\@parameterinit{2}}%
     \def\Bigfonts{\xivpt}\def\Bigsize{\xivpt\@parameterinit{4}}%
     \def\biggfonts{\xviipt}\def\biggsize{\xviipt\@parameterinit{6}}%
     \def\Biggfonts{\xxipt}\def\Biggsize{\xxipt\@parameterinit{8}}%
     \def\tinyfonts{\vpt}\def\tinysize{\vpt\@parameterinit{-8}}%
     \def\HUGEFONTS{\xxvpt}\def\HUGESIZE{\xxvpt\@parameterinit{10}}%
     \normalsize\fixedskipslist}

\def\@setsizexii{\getxipt
     \def\subsubscriptfonts{\vipt}%
          \def\subsubscriptsize{\vipt\@parameterinit{-6}}%
     \def\subscriptfonts{\viiipt}%
          \def\subscriptsize{\viiipt\@parameterinit{-2}}%
     \def\footnotefonts{\xpt}\def\footnotesize{\xpt\@parameterinit{0}}%
     \def\smallfonts{\xipt}\def\smallsize{\xipt\@parameterinit{1}}%
     \def\normalfonts{\xiipt}\def\normalsize{\xiipt\@parameterinit{2}}%
     \def\bigfonts{\xivpt}\def\bigsize{\xivpt\@parameterinit{4}}%
     \def\Bigfonts{\xviipt}\def\Bigsize{\xviipt\@parameterinit{6}}%
     \def\biggfonts{\xxipt}\def\biggsize{\xxipt\@parameterinit{8}}%
     \def\Biggfonts{\xxvpt}\def\Biggsize{\xxvpt\@parameterinit{10}}%
     \def\tinyfonts{\vpt}\def\tinysize{\vpt\@parameterinit{-8}}%
     \def\HUGEFONTS{\xxvpt}\def\HUGESIZE{\xxvpt\@parameterinit{10}}%
     \normalsize\fixedskipslist}

\def\@setsizexiv{\getxiiipt
     \def\subsubscriptfonts{\viipt}%
          \def\subsubscriptsize{\viipt\@parameterinit{-4}}%
     \def\subscriptfonts{\xpt}\def\subscriptsize{\xpt\@parameterinit{0}}%
     \def\footnotefonts{\xiipt}\def\footnotesize{\xiipt\@parameterinit{2}}%
     \def\smallfonts{\xiiipt}\def\smallsize{\xiiipt\@parameterinit{3}}%
     \def\normalfonts{\xivpt}\def\normalsize{\xivpt\@parameterinit{4}}%
     \def\bigfonts{\xviipt}\def\bigsize{\xviipt\@parameterinit{6}}%
     \def\Bigfonts{\xxipt}\def\Bigsize{\xxipt\@parameterinit{8}}%
     \def\biggfonts{\xxvpt}\def\biggsize{\xxvpt\@parameterinit{10}}%
     \def\Biggfonts{\err@sizetoolarge\Biggfonts\HUGEFONTS}%
          \def\Biggsize{\err@sizetoolarge\Biggsize\HUGESIZE}%
     \def\tinyfonts{\vpt}\def\tinysize{\vpt\@parameterinit{-8}}%
     \def\HUGEFONTS{\xxvpt}\def\HUGESIZE{\xxvpt\@parameterinit{10}}%
     \normalsize\fixedskipslist}

\def\subsubscriptfonts{\vpt} \def\subsubscriptsize{\vpt\@parameterinit{-8}}
\def\subscriptfonts{\viipt}  \def\subscriptsize{\viipt\@parameterinit{-4}}
\def\footnotefonts{\viiipt}  \def\footnotesize{\viiipt\@parameterinit{-2}}
\def\smallfonts{\err@sizenotavailable\smallfonts}
                             \def\smallsize{\ixpt\@parameterinit{-1}}
\def\normalfonts{\xpt}       \def\normalsize{\xpt\@parameterinit{0}}
\def\bigfonts{\xiipt}        \def\bigsize{\xiipt\@parameterinit{2}}
\def\Bigfonts{\xivpt}        \def\Bigsize{\xivpt\@parameterinit{4}}
\def\biggfonts{\xviipt}      \def\biggsize{\xviipt\@parameterinit{6}}
\def\Biggfonts{\xxipt}       \def\Biggsize{\xxipt\@parameterinit{8}}
\def\tinyfonts{\vpt}         \def\tinysize{\vpt\@parameterinit{-8}}
\def\HUGEFONTS{\xxvpt}       \def\HUGESIZE{\xxvpt\@parameterinit{10}}

\message{document layout,}


\newtoks\everyoutput \everyoutput={}
\newdimen\depthofpage
\newcount\pagenum \pagenum=0

\newdimen\oddtopmargin  \newdimen\eventopmargin
\newdimen\oddleftmargin \newdimen\evenleftmargin
\newtoks\oddhead        \newtoks\evenhead
\newtoks\oddfoot        \newtoks\evenfoot

\def\topmargin{\afterassignment\@seteventop\oddtopmargin}
\def\leftmargin{\afterassignment\@setevenleft\oddleftmargin}
\def\head{\afterassignment\@setevenhead\oddhead}
\def\foot{\afterassignment\@setevenfoot\oddfoot}

\def\@seteventop{\eventopmargin=\oddtopmargin}
\def\@setevenleft{\evenleftmargin=\oddleftmargin}
\def\@setevenhead{\evenhead=\oddhead}
\def\@setevenfoot{\evenfoot=\oddfoot}

\def\pagenumstyle#1{\@setnumstyle\pagenum{#1}}

\newif\ifdraft
\def\draft{\drafttrue\leftmargin=.5in \overfullrule=5pt }

\def\outputstyle#1{\global\expandafter\let\expandafter
          \@outputstyle\csname#1output\endcsname
     \usename{#1setup}}

\output={\@outputstyle}

\def\normaloutput{\the\everyoutput
     \global\advance\pagenum by\@ne
     \ifodd\pagenum
          \voffset=\oddtopmargin \hoffset=\oddleftmargin
     \else\voffset=\eventopmargin \hoffset=\evenleftmargin
     \fi
     \advance\voffset by-1in  \advance\hoffset by-1in
     \count0=\pagenum
     \expandafter\shipout\pagebox
     \ifnum\outputpenalty>-\@MM\else\dosupereject\fi}

\newdimen\fullhsize
\newbox\leftpage
\newcount\leftpagenum
\newcount\outputpagenum \outputpagenum=0
\let\leftorright=L

\def\twoupoutput{\the\everyoutput
     \global\advance\pagenum by\@ne
     \if L\leftorright
          \global\setbox\leftpage=\leftline{\pagebox}%
          \global\leftpagenum=\pagenum
          \global\let\leftorright=R%
     \else\global\advance\outputpagenum by\@ne
          \ifodd\outputpagenum
               \voffset=\oddtopmargin \hoffset=\oddleftmargin
          \else\voffset=\eventopmargin \hoffset=\evenleftmargin
          \fi
          \advance\voffset by-1in  \advance\hoffset by-1in
          \count0=\leftpagenum \count1=\pagenum
          \shipout\vbox{\hbox to\fullhsize
               {\box\leftpage\hfil\leftline{\pagebox}}}%
          \global\let\leftorright=L%
     \fi
     \ifnum\outputpenalty>-\@MM
     \else\dosupereject
          \if R\leftorright
               \globaldefs=\@ne\head={\hfil}\foot={\hfil}\globaldefs=\z@
               \null\newpage
          \fi
     \fi}

\def\pagebox{\vbox{\makeheadline\pagebody\makefootline}}

\def\makeheadline{%
     \vbox to\z@{\baselinestretch=\@m
          \vskip\topskip\vskip-.708\baselineskip\vskip-\headskip
          \line{\vbox to\ht\strutbox{}%
               \ifodd\pagenum\the\oddhead\else\the\evenhead\fi}%
          \vss}%
     \nointerlineskip}

\def\pagebody{\vbox to\vsize{%
     \boxmaxdepth\maxdepth
     \ifvoid\topins\else\unvbox\topins\fi
     \depthofpage=\dp255
     \unvbox255
     \ifraggedbottom\kern-\depthofpage\vfil\fi
     \ifvoid\footins
     \else\vskip\skip\footins
          \footnoterule
          \unvbox\footins
          \vskip-\footnoteskip
     \fi}}

\def\makefootline{\baselineskip=\footskip
     \line{\ifodd\pagenum\the\oddfoot\else\the\evenfoot\fi}}


\newskip\abovechapterskip
\newskip\belowchapterskip
\newskip\abovesectionskip
\newskip\belowsectionskip
\newskip\abovesubsectionskip
\newskip\belowsubsectionskip

\def\chapterstyle#1{\global\expandafter\let\expandafter\@chapterstyle
     \csname#1text\endcsname}
\def\sectionstyle#1{\global\expandafter\let\expandafter\@sectionstyle
     \csname#1text\endcsname}
\def\subsectionstyle#1{\global\expandafter\let\expandafter\@subsectionstyle
     \csname#1text\endcsname}

\def\chapter#1{%
     \ifdim\lastskip=17sp \else\chapterbreak\vskip\abovechapterskip\fi
     \@chapterstyle{\ifblank\chapternumstyle\then
          \else\newchapternum=\next\chapternumformat\ \fi#1}%
     \nobreak\vskip\belowchapterskip\vskip17sp }

\def\section#1{%
     \ifdim\lastskip=17sp \else\sectionbreak\vskip\abovesectionskip\fi
     \@sectionstyle{\ifblank\sectionnumstyle\then
          \else\newsectionnum=\next\sectionnumformat\ \fi#1}%
     \nobreak\vskip\belowsectionskip\vskip17sp }

\def\subsection#1{%
     \ifdim\lastskip=17sp \else\subsectionbreak\vskip\abovesubsectionskip\fi
     \@subsectionstyle{\ifblank\subsectionnumstyle\then
          \else\newsubsectionnum=\next\subsectionnumformat\ \fi#1}%
     \nobreak\vskip\belowsubsectionskip\vskip17sp }


\let\TeXunderline=\underline
\let\TeXoverline=\overline
\def\underline#1{\relax\ifmmode\TeXunderline{#1}\else
     $\TeXunderline{\hbox{#1}}$\fi}
\def\overline#1{\relax\ifmmode\TeXoverline{#1}\else
     $\TeXoverline{\hbox{#1}}$\fi}

\def\baselinestretch{\afterassignment\@baselinestretch\count@}
\def\@baselinestretch{\baselineskip=\normalbaselineskip
     \divide\baselineskip by\@m\baselineskip=\count@\baselineskip
     \setbox\strutbox=\hbox{\vrule
          height.708\baselineskip depth.292\baselineskip width\z@}%
     \bigskipamount=\the\baselineskip
          plus.25\baselineskip minus.25\baselineskip
     \medskipamount=.5\baselineskip
          plus.125\baselineskip minus.125\baselineskip
     \smallskipamount=.25\baselineskip
          plus.0625\baselineskip minus.0625\baselineskip}

\def\\{\ifhmode\ifnum\lastpenalty=-\@M\else\hfil\penalty-\@M\fi\fi
     \ignorespaces}
\def\newpage{\vfil\break}

\def\lefttext#1{\par{\@text\leftskip=\z@\rightskip=\centering
     \noindent#1\par}}
\def\righttext#1{\par{\@text\leftskip=\centering\rightskip=\z@
     \noindent#1\par}}
\def\centertext#1{\par{\@text\leftskip=\centering\rightskip=\centering
     \noindent#1\par}}
\def\@text{\parindent=\z@ \parfillskip=\z@ \everypar={}%
     \spaceskip=.3333em \xspaceskip=.5em
     \def\\{\ifhmode\ifnum\lastpenalty=-\@M\else\penalty-\@M\fi\fi
          \ignorespaces}}

\def\beginleft{\par\@text\leftskip=\z@ \rightskip=\centering}
     
\def\beginright{\par\@text\leftskip=\centering\rightskip=\z@ }
     
\def\begincenter{\par\@text\leftskip=\centering\rightskip=\centering}

\def\beginnarrow{\defaultoption[\parindent]\@beginnarrow}
\def\@beginnarrow[#1]{\par\advance\leftskip by#1\advance\rightskip by#1}

\begingroup
\catcode`\[=1 \catcode`\{=11
\gdef\beginignore[\endgroup\bgroup
     \catcode`\e=0 \catcode`\\=12 \catcode`\{=11 \catcode`\f=12 \let\or=\relax
     \let\nd{ignor=\fi \let\}=\egroup
     \iffalse}
\endgroup

\long\def\marginnote#1{\leavevmode
     \edef\@marginsf{\spacefactor=\the\spacefactor\relax}%
     \ifdraft\strut\vadjust{%
          \hbox to\z@{\hskip\hsize\hskip.1in
               \vbox to\z@{\vskip-\dp\strutbox
                    \marginnoteformat
                    \vskip-\ht\strutbox
                    \noindent\strut#1\par
                    \vss}%
               \hss}}%
     \fi
     \@marginsf}


\newtoks\everybye \everybye={\par\vfil}
\outer\def\bye{\the\everybye
     \footnotecheck
     \prelabelcheck
     \streamcheck
     \supereject
     \TeXend}

\message{footnotes,}

\newcount\footnotenum \footnotenum=0
\newskip\footnoteskip
\let\@footnotelist=\empty

\def\footnotenumstyle#1{\@setnumstyle\footnotenum{#1}%
     \useafter\ifx{@footnotenumstyle}\symbols
          \global\let\@footup=\empty
     \else\global\let\@footup=\markup
     \fi}

\def\footnote{\footnotecheck\defaultoption[]\@footnote}
\def\@footnote[#1]{\@footnotemark[#1]\@footnotetext}

\def\footnotemark{\defaultoption[]\@footnotemark}
\def\@footnotemark[#1]{\let\@footsf=\empty
     \ifhmode\edef\@footsf{\spacefactor=\the\spacefactor\relax}\/\fi
     \ifnoarg#1\then
          \global\advance\footnotenum by\@ne
          \@footup{\footnotenumformat}%
          \edef\@@foota{\footnotenum=\the\footnotenum\relax}%
          \expandafter\additemR\expandafter\@footup\expandafter
               {\@@foota\footnotenumformat}\to\@footnotelist
          \global\let\@footnotelist=\@footnotelist
     \else\markup{#1}%
          \additemR\markup{#1}\to\@footnotelist
          \global\let\@footnotelist=\@footnotelist
     \fi
     \@footsf}

\def\footnotetext{%
     \ifx\@footnotelist\empty\err@extrafootnotetext\else\@footnotetext\fi}
\def\@footnotetext{%
     \getitemL\@footnotelist\to\@@foota
     \global\let\@footnotelist=\@footnotelist
     \insert\footins\bgroup
     \footnoteformat
     \splittopskip=\ht\strutbox\splitmaxdepth=\dp\strutbox
     \interlinepenalty=\interfootnotelinepenalty\floatingpenalty=\@MM
     \noindent\llap{\@@foota}\strut
     \bgroup\aftergroup\@footnoteend
     \let\@@scratcha=}
\def\@footnoteend{\strut\par\vskip\footnoteskip\egroup}

\def\footnoterule{\normalfonts
     \kern-.3em \hrule width2in height.04em \kern .26em }

\def\footnotecheck{%
     \ifx\@footnotelist\empty
     \else\err@extrafootnotemark
          \global\let\@footnotelist=\empty
     \fi}

\message{labels,}

\let\@@labeldef=\xdef
\newif\if@labelfile
\newwrite\@labelfile
\let\@prelabellist=\empty

\def\label#1#2{\trim#1\to\@@labarg\edef\@@labtext{#2}%
     \edef\@@labname{lab@\@@labarg}%
     \useafter\ifundefined\@@labname\then\else\@yeslab\fi
     \useafter\@@labeldef\@@labname{#2}%
     \ifstreaming
          \expandafter\toks@\expandafter\expandafter\expandafter
               {\csname\@@labname\endcsname}%
          \immediate\write\streamout{\noexpand\label{\@@labarg}{\the\toks@}}%
     \fi}
\def\@yeslab{%
     \useafter\ifundefined{if\@@labname}\then
          \err@labelredef\@@labarg
     \else\useif{if\@@labname}\then
               \err@labelredef\@@labarg
          \else\global\usename{\@@labname true}%
               \useafter\ifundefined{pre\@@labname}\then
               \else\useafter\ifx{pre\@@labname}\@@labtext
                    \else\err@badlabelmatch\@@labarg
                    \fi
               \fi
               \if@labelfile
               \else\global\@labelfiletrue
                    \immediate\write\sixt@@n{--> Creating file \jobname.lab}%
                    \immediate\openout\@labelfile=\jobname.lab
               \fi
               \immediate\write\@labelfile
                    {\noexpand\prelabel{\@@labarg}{\@@labtext}}%
          \fi
     \fi}

\def\putlab#1{\trim#1\to\@@labarg\edef\@@labname{lab@\@@labarg}%
     \useafter\ifundefined\@@labname\then\@nolab\else\usename\@@labname\fi}
\def\@nolab{%
     \useafter\ifundefined{pre\@@labname}\then
          \undefinedlabelformat
          \err@needlabel\@@labarg
          \useafter\xdef\@@labname{\undefinedlabelformat}%
     \else\usename{pre\@@labname}%
          \useafter\xdef\@@labname{\usename{pre\@@labname}}%
     \fi
     \useafter\newif{if\@@labname}%
     \expandafter\additemR\@@labarg\to\@prelabellist}

\def\prelabel#1{\useafter\gdef{prelab@#1}}

\def\ifundefinedlabel#1\then{%
     \expandafter\ifx\csname lab@#1\endcsname\relax}
\def\useiflab#1\then{\csname iflab@#1\endcsname}

\def\prelabelcheck{{%
     \def\^^\##1{\useiflab{##1}\then\else\err@undefinedlabel{##1}\fi}%
     \@prelabellist}}

\message{equation numbering,}

\newcount\chapternum
\newcount\sectionnum
\newcount\subsectionnum
\newcount\equationnum
\newcount\subequationnum
\newcount\figurenum
\newcount\subfigurenum
\newcount\tablenum
\newcount\subtablenum

\newif\if@subeqncount
\newif\if@subfigcount
\newif\if@subtblcount

\def\newchapternum{\newsectionnum=\z@\@resetnum\chapternum}
\def\newsectionnum{\newsubsectionnum=\z@\@resetnum\sectionnum}
\def\newsubsectionnum{\newequationnum=\z@\newfigurenum=\z@\newtablenum=\z@
     \@resetnum\subsectionnum}
\def\newequationnum{\newsubequationnum=\z@\@resetnum\equationnum}
\def\newsubequationnum{\@resetnum\subequationnum}
\def\newfigurenum{\newsubfigurenum=\z@\@resetnum\figurenum}
\def\newsubfigurenum{\@resetnum\subfigurenum}
\def\newtablenum{\newsubtablenum=\z@\@resetnum\tablenum}
\def\newsubtablenum{\@resetnum\subtablenum}

\def\@resetnum#1{\global\advance#1by1 \edef\next{\the#1\relax}\global#1}

\newchapternum=0

\def\chapternumstyle#1{\@setnumstyle\chapternum{#1}}
\def\sectionnumstyle#1{\@setnumstyle\sectionnum{#1}}
\def\subsectionnumstyle#1{\@setnumstyle\subsectionnum{#1}}
\def\equationnumstyle#1{\@setnumstyle\equationnum{#1}}
\def\subequationnumstyle#1{\@setnumstyle\subequationnum{#1}%
     \ifblank\subequationnumstyle\then\global\@subeqncountfalse\fi
     \ignorespaces}
\def\figurenumstyle#1{\@setnumstyle\figurenum{#1}}
\def\subfigurenumstyle#1{\@setnumstyle\subfigurenum{#1}%
     \ifblank\subfigurenumstyle\then\global\@subfigcountfalse\fi
     \ignorespaces}
\def\tablenumstyle#1{\@setnumstyle\tablenum{#1}}
\def\subtablenumstyle#1{\@setnumstyle\subtablenum{#1}%
     \ifblank\subtablenumstyle\then\global\@subtblcountfalse\fi
     \ignorespaces}

\def\eqnlabel#1{%
     \if@subeqncount
          \newsubequationnum=\next
     \else\newequationnum=\next
          \ifblank\subequationnumstyle\then
          \else\global\@subeqncounttrue
               \newsubequationnum=\@ne
          \fi
     \fi
     \label{#1}{\puteqnformat}(\puteqn{#1})%
     \ifdraft\rlap{\hskip.1in{\tt#1}}\fi}

\let\puteqn=\putlab

\def\equation#1#2{\useafter\gdef{eqn@#1}{#2\eqno\eqnlabel{#1}}}
\def\Equation#1{\useafter\gdef{eqn@#1}}

\def\putequation#1{\useafter\ifundefined{eqn@#1}\then
     \err@undefinedeqn{#1}\else\usename{eqn@#1}\fi}

\def\eqnseriesstyle#1{\gdef\@eqnseriesstyle{#1}}
\def\begineqnseries{\subequationnumstyle{\@eqnseriesstyle}%
     \defaultoption[]\@begineqnseries}
\def\@begineqnseries[#1]{\edef\@@eqnname{#1}}
\def\endeqnseries{\subequationnumstyle{blank}%
     \expandafter\ifnoarg\@@eqnname\then
     \else\label\@@eqnname{\puteqnformat}%
     \fi
     \aftergroup\ignorespaces}

\def\figlabel#1{%
     \if@subfigcount
          \newsubfigurenum=\next
     \else\newfigurenum=\next
          \ifblank\subfigurenumstyle\then
          \else\global\@subfigcounttrue
               \newsubfigurenum=\@ne
          \fi
     \fi
     \label{#1}{\putfigformat}\putfig{#1}%
     {\def\marginnoteformat{\tt}\marginnote{#1}}}

\let\putfig=\putlab

\def\figseriesstyle#1{\gdef\@figseriesstyle{#1}}
\def\beginfigseries{\subfigurenumstyle{\@figseriesstyle}%
     \defaultoption[]\@beginfigseries}
\def\@beginfigseries[#1]{\edef\@@figname{#1}}
\def\endfigseries{\subfigurenumstyle{blank}%
     \expandafter\ifnoarg\@@figname\then
     \else\label\@@figname{\putfigformat}%
     \fi
     \aftergroup\ignorespaces}

\def\tbllabel#1{%
     \if@subtblcount
          \newsubtablenum=\next
     \else\newtablenum=\next
          \ifblank\subtablenumstyle\then
          \else\global\@subtblcounttrue
               \newsubtablenum=\@ne
          \fi
     \fi
     \label{#1}{\puttblformat}\puttbl{#1}%
     {\def\marginnoteformat{\tt}\marginnote{#1}}}

\let\puttbl=\putlab

\def\tblseriesstyle#1{\gdef\@tblseriesstyle{#1}}
\def\begintblseries{\subtablenumstyle{\@tblseriesstyle}%
     \defaultoption[]\@begintblseries}
\def\@begintblseries[#1]{\edef\@@tblname{#1}}
\def\endtblseries{\subtablenumstyle{blank}%
     \expandafter\ifnoarg\@@tblname\then
     \else\label\@@tblname{\puttblformat}%
     \fi
     \aftergroup\ignorespaces}

\message{reference numbering,}

\newcount\referencenum \referencenum=0
\newcount\@@prerefcount \@@prerefcount=0
\newcount\@@thisref
\newcount\@@lastref
\newcount\@@loopref
\newcount\@@refseq
\newdimen\refnumindent
\let\@undefreflist=\empty

\def\referencenumstyle#1{\@setnumstyle\referencenum{#1}}

\def\referencestyle#1{\usename{@ref#1}}

\def\@refsequential{%
     \gdef\@refpredef##1{\global\advance\referencenum by\@ne
          \let\^^\=0\label{##1}{\^^\{\the\referencenum}}%
          \useafter\gdef{ref@\the\referencenum}{{##1}{\undefinedlabelformat}}}%
     \gdef\@reference##1##2{%
          \ifundefinedlabel##1\then
          \else\def\^^\####1{\global\@@thisref=####1\relax}\putlab{##1}%
               \useafter\gdef{ref@\the\@@thisref}{{##1}{##2}}%
          \fi}%
     \gdef\endputreferences{%
          \loop\ifnum\@@loopref<\referencenum
                    \advance\@@loopref by\@ne
                    \expandafter\expandafter\expandafter\@printreference
                         \csname ref@\the\@@loopref\endcsname
          \repeat
          \par}}

\def\@refpreordered{%
     \gdef\@refpredef##1{\global\advance\referencenum by\@ne
          \additemR##1\to\@undefreflist}%
     \gdef\@reference##1##2{%
          \ifundefinedlabel##1\then
          \else\global\advance\@@loopref by\@ne
               {\let\^^\=0\label{##1}{\^^\{\the\@@loopref}}}%
               \@printreference{##1}{##2}%
          \fi}
     \gdef\endputreferences{%
          \def\^^\####1{\useiflab{####1}\then
               \else\reference{####1}{\undefinedlabelformat}\fi}%
          \@undefreflist
          \par}}

\def\beginprereferences{\par
     \def\reference##1##2{\global\advance\referencenum by1\@ne
          \let\^^\=0\label{##1}{\^^\{\the\referencenum}}%
          \useafter\gdef{ref@\the\referencenum}{{##1}{##2}}}}
\def\endprereferences{\global\@@prerefcount=\the\referencenum\par}

\def\beginputreferences{\par
     \refnumindent=\z@\@@loopref=\z@
     \loop\ifnum\@@loopref<\referencenum
               \advance\@@loopref by\@ne
               \setbox\z@=\hbox{\referencenum=\@@loopref
                    \referencenumformat\enskip}%
               \ifdim\wd\z@>\refnumindent\refnumindent=\wd\z@\fi
     \repeat
     \putreferenceformat
     \@@loopref=\z@
     \loop\ifnum\@@loopref<\@@prerefcount
               \advance\@@loopref by\@ne
               \expandafter\expandafter\expandafter\@printreference
                    \csname ref@\the\@@loopref\endcsname
     \repeat
     \let\reference=\@reference}

\def\@printreference#1#2{\ifx#2\undefinedlabelformat\err@undefinedref{#1}\fi
     \noindent\ifdraft\rlap{\hskip\hsize\hskip.1in \tt#1}\fi
     \llap{\referencenum=\@@loopref\referencenumformat\enskip}#2\par}

\def\reference#1#2{{\par\refnumindent=\z@\putreferenceformat\noindent#2\par}}

\def\putref#1{\trim#1\to\@@refarg
     \expandafter\ifnoarg\@@refarg\then
          \toks@={\relax}%
     \else\@@lastref=-\@m\def\@@refsep{}\def\@more{\@nextref}%
          \toks@={\@nextref#1,,}%
     \fi\the\toks@}
\def\@nextref#1,{\trim#1\to\@@refarg
     \expandafter\ifnoarg\@@refarg\then
          \let\@more=\relax
     \else\ifundefinedlabel\@@refarg\then
               \expandafter\@refpredef\expandafter{\@@refarg}%
          \fi
          \def\^^\##1{\global\@@thisref=##1\relax}%
          \global\@@thisref=\m@ne
          \setbox\z@=\hbox{\putlab\@@refarg}%
     \fi
     \advance\@@lastref by\@ne
     \ifnum\@@lastref=\@@thisref\advance\@@refseq by\@ne\else\@@refseq=\@ne\fi
     \ifnum\@@lastref<\z@
     \else\ifnum\@@refseq<\thr@@
               \@@refsep\def\@@refsep{,}%
               \ifnum\@@lastref>\z@
                    \advance\@@lastref by\m@ne
                    {\referencenum=\@@lastref\putrefformat}%
               \else\undefinedlabelformat
               \fi
          \else\def\@@refsep{--}%
          \fi
     \fi
     \@@lastref=\@@thisref
     \@more}

\message{streaming,}

\newif\ifstreaming

\def\streamto{\defaultoption[\jobname]\@streamto}
\def\@streamto[#1]{\global\streamingtrue
     \immediate\write\sixt@@n{--> Streaming to #1.str}%
     \newwrite\streamout\immediate\openout\streamout=#1.str }

\def\streamfrom{\defaultoption[\jobname]\@streamfrom}
\def\@streamfrom[#1]{\newread\streamin\openin\streamin=#1.str
     \ifeof\streamin
          \expandafter\err@nostream\expandafter{#1.str}%
     \else\immediate\write\sixt@@n{--> Streaming from #1.str}%
          \let\@@labeldef=\gdef
          \ifstreaming
               \edef\@elc{\endlinechar=\the\endlinechar}%
               \endlinechar=\m@ne
               \loop\read\streamin to\@@scratcha
                    \ifeof\streamin
                         \streamingfalse
                    \else\toks@=\expandafter{\@@scratcha}%
                         \immediate\write\streamout{\the\toks@}%
                    \fi
                    \ifstreaming
               \repeat
               \@elc
               \input #1.str
               \streamingtrue
          \else\input #1.str
          \fi
          \let\@@labeldef=\xdef
     \fi}

\def\streamcheck{\ifstreaming
     \immediate\write\streamout{\pagenum=\the\pagenum}%
     \immediate\write\streamout{\footnotenum=\the\footnotenum}%
     \immediate\write\streamout{\referencenum=\the\referencenum}%
     \immediate\write\streamout{\chapternum=\the\chapternum}%
     \immediate\write\streamout{\sectionnum=\the\sectionnum}%
     \immediate\write\streamout{\subsectionnum=\the\subsectionnum}%
     \immediate\write\streamout{\equationnum=\the\equationnum}%
     \immediate\write\streamout{\subequationnum=\the\subequationnum}%
     \immediate\write\streamout{\figurenum=\the\figurenum}%
     \immediate\write\streamout{\subfigurenum=\the\subfigurenum}%
     \immediate\write\streamout{\tablenum=\the\tablenum}%
     \immediate\write\streamout{\subtablenum=\the\subtablenum}%
     \immediate\closeout\streamout
     \fi}


\def\err@badtypesize{%
     \errhelp={The limited availability of certain fonts requires^^J%
          that the base type size be 10pt, 12pt, or 14pt.^^J}%
     \errmessage{--> Illegal base type size}}

\def\err@badsizechange{\immediate\write\sixt@@n
     {--> Size change not allowed in math mode, ignored}}

\def\err@sizetoolarge#1{\immediate\write\sixt@@n
     {--> \noexpand#1 too big, substituting HUGE}}

\def\err@sizenotavailable#1{\immediate\write\sixt@@n
     {--> Size not available, \noexpand#1 ignored}}

\def\err@fontnotavailable#1{\immediate\write\sixt@@n
     {--> Font not available, \noexpand#1 ignored}}

\def\err@sltoit{\immediate\write\sixt@@n
     {--> Style \noexpand\sl not available, substituting \noexpand\it}%
     \it}

\def\err@bfstobf{\immediate\write\sixt@@n
     {--> Style \noexpand\bfs not available, substituting \noexpand\bf}%
     \bf}

\def\err@badgroup#1#2{%
     \errhelp={The block you have just tried to close was not the one^^J%
          most recently opened.^^J}%
     \errmessage{--> \noexpand\end{#1} doesn't match \noexpand\begin{#2}}}

\def\err@badcountervalue#1{\immediate\write\sixt@@n
     {--> Counter (#1) out of bounds}}

\def\err@extrafootnotemark{\immediate\write\sixt@@n
     {--> \noexpand\footnotemark command
          has no corresponding \noexpand\footnotetext}}

\def\err@extrafootnotetext{%
     \errhelp{You have given a \noexpand\footnotetext command without first
          specifying^^Ja \noexpand\footnotemark.^^J}%
     \errmessage{--> \noexpand\footnotetext command has no corresponding
          \noexpand\footnotemark}}

\def\err@labelredef#1{\immediate\write\sixt@@n
     {--> Label "#1" redefined}}

\def\err@badlabelmatch#1{\immediate\write\sixt@@n
     {--> Definition of label "#1" doesn't match value in \jobname.lab}}

\def\err@needlabel#1{\immediate\write\sixt@@n
     {--> Label "#1" cited before its definition}}

\def\err@undefinedlabel#1{\immediate\write\sixt@@n
     {--> Label "#1" cited but never defined}}

\def\err@undefinedeqn#1{\immediate\write\sixt@@n
     {--> Equation "#1" not defined}}

\def\err@undefinedref#1{\immediate\write\sixt@@n
     {--> Reference "#1" not defined}}

\def\err@nostream#1{%
     \errhelp={You have tried to input a stream file that doesn't exist.^^J}%
     \errmessage{--> Stream file #1 not found}}

\message{jyTeX initialization}

\everyjob{\immediate\write16{--> jyTeX version \fmtversion}%
     \edef\@@jobname{\jobname}%
     \edef\jobname{\@@jobname}%
     \settime
     \openin0=\jobname.lab
     \ifeof0
     \else\closein0
          \immediate\write16{--> Getting labels from file \jobname.lab}%
          \input\jobname.lab
     \fi}


\def\fixedskipslist{%
     \^^\{\topskip}%
     \^^\{\splittopskip}%
     \^^\{\maxdepth}%
     \^^\{\skip\topins}%
     \^^\{\skip\footins}%
     \^^\{\headskip}%
     \^^\{\footskip}}

\def\scalingskipslist{%
     \^^\{\p@renwd}%
     \^^\{\delimitershortfall}%
     \^^\{\nulldelimiterspace}%
     \^^\{\scriptspace}%
     \^^\{\jot}%
     \^^\{\normalbaselineskip}%
     \^^\{\normallineskip}%
     \^^\{\normallineskiplimit}%
     \^^\{\baselineskip}%
     \^^\{\lineskip}%
     \^^\{\lineskiplimit}%
     \^^\{\bigskipamount}%
     \^^\{\medskipamount}%
     \^^\{\smallskipamount}%
     \^^\{\parskip}%
     \^^\{\parindent}%
     \^^\{\abovedisplayskip}%
     \^^\{\belowdisplayskip}%
     \^^\{\abovedisplayshortskip}%
     \^^\{\belowdisplayshortskip}%
     \^^\{\abovechapterskip}%
     \^^\{\belowchapterskip}%
     \^^\{\abovesectionskip}%
     \^^\{\belowsectionskip}%
     \^^\{\abovesubsectionskip}%
     \^^\{\belowsubsectionskip}}


\def\twoupsetup{
     \topmargin=.75in
     \leftmargin=.5in
     \vsize=6.9in
     \hsize=4.75in
     \fullhsize=10in
     \let\draft=\relax}

\outputstyle{normal}                             

\def\marginnoteformat{\subscriptsize             
     \hsize=1in \baselinestretch=1000 \everypar={}%
     \tolerance=5000 \hbadness=5000 \parskip=0pt \parindent=0pt
     \leftskip=0pt \rightskip=0pt \raggedright}

\head={\ifdraft\normalfonts\it\hfil DRAFT\hfil   
     \llap{\number\day\ \monthword\month\ \militarytime}\else\hfil\fi}
\foot={\hfil\normalfonts\numstyle\pagenum\hfil}  

\normalbaselineskip=12pt                         
\normallineskip=0pt                              
\normallineskiplimit=0pt                         
\normalbaselines                                 

\topskip=.85\baselineskip
\splittopskip=\topskip
\headskip=2\baselineskip
\footskip=\headskip

\pagenumstyle{arabic}                            

\parskip=0pt                                     
\parindent=20pt                                  

\baselinestretch=1000                            


\chapterstyle{left}                              
\chapternumstyle{blank}                          
\def\chapterbreak{\newpage}                      
\abovechapterskip=0pt                            
\belowchapterskip=1.5\baselineskip               
     plus.38\baselineskip minus.38\baselineskip
\def\chapternumformat{\numstyle\chapternum.}     

\sectionstyle{left}                              
\sectionnumstyle{blank}                          
\def\sectionbreak{\vskip0pt plus4\baselineskip\penalty-100
     \vskip0pt plus-4\baselineskip}              
\abovesectionskip=1.5\baselineskip               
     plus.38\baselineskip minus.38\baselineskip
\belowsectionskip=\the\baselineskip              
     plus.25\baselineskip minus.25\baselineskip
\def\sectionnumformat{
     \ifblank\chapternumstyle\then\else\numstyle\chapternum.\fi
     \numstyle\sectionnum.}

\subsectionstyle{left}                           
\subsectionnumstyle{blank}                       
\def\subsectionbreak{\vskip0pt plus4\baselineskip\penalty-100
     \vskip0pt plus-4\baselineskip}              
\abovesubsectionskip=\the\baselineskip           
     plus.25\baselineskip minus.25\baselineskip
\belowsubsectionskip=.75\baselineskip            
     plus.19\baselineskip minus.19\baselineskip
\def\subsectionnumformat{
     \ifblank\chapternumstyle\then\else\numstyle\chapternum.\fi
     \ifblank\sectionnumstyle\then\else\numstyle\sectionnum.\fi
     \numstyle\subsectionnum.}


\footnotenumstyle{symbols}                       
\footnoteskip=0pt                                
\def\footnotenumformat{\numstyle\footnotenum}    
\def\footnoteformat{\footnotesize                
     \everypar={}\parskip=0pt \parfillskip=0pt plus1fil
     \leftskip=1em \rightskip=0pt
     \spaceskip=0pt \xspaceskip=0pt
     \def\\{\ifhmode\ifnum\lastpenalty=-10000
          \else\hfil\penalty-10000 \fi\fi\ignorespaces}}


\def\undefinedlabelformat{$\bullet$}             


\equationnumstyle{arabic}                        
\subequationnumstyle{blank}                      
\figurenumstyle{arabic}                          
\subfigurenumstyle{blank}                        
\tablenumstyle{arabic}                           
\subtablenumstyle{blank}                         

\eqnseriesstyle{alphabetic}                      
\figseriesstyle{alphabetic}                      
\tblseriesstyle{alphabetic}                      

\def\puteqnformat{\hbox{
     \ifblank\chapternumstyle\then\else\numstyle\chapternum.\fi
     \ifblank\sectionnumstyle\then\else\numstyle\sectionnum.\fi
     \ifblank\subsectionnumstyle\then\else\numstyle\subsectionnum.\fi
     \numstyle\equationnum
     \numstyle\subequationnum}}
\def\putfigformat{\hbox{
     \ifblank\chapternumstyle\then\else\numstyle\chapternum.\fi
     \ifblank\sectionnumstyle\then\else\numstyle\sectionnum.\fi
     \ifblank\subsectionnumstyle\then\else\numstyle\subsectionnum.\fi
     \numstyle\figurenum
     \numstyle\subfigurenum}}
\def\puttblformat{\hbox{
     \ifblank\chapternumstyle\then\else\numstyle\chapternum.\fi
     \ifblank\sectionnumstyle\then\else\numstyle\sectionnum.\fi
     \ifblank\subsectionnumstyle\then\else\numstyle\subsectionnum.\fi
     \numstyle\tablenum
     \numstyle\subtablenum}}


\referencestyle{sequential}                      
\referencenumstyle{arabic}                       
\def\putrefformat{\numstyle\referencenum}        
\def\referencenumformat{\numstyle\referencenum.} 
\def\putreferenceformat{
     \everypar={\hangindent=1em \hangafter=1 }%
     \def\\{\hfil\break\null\hskip-1em \ignorespaces}%
     \leftskip=\refnumindent\parindent=0pt \interlinepenalty=1000 }


\normalsize


\def\fmtversion{2.6M (June 1992)}

\catcode`\@=12
\def\upref#1/{\markup{[\putref{#1}]}}

\typesize=12pt

\footnoteskip=2pt
\footnotenumstyle{arabic}
%
%
\def\JSP#1#2#3{{\sl J. Stat. Phys.} {\bf #1} (#2) #3}
\def\PRL#1#2#3{{\sl Phys. Rev. Lett.} {\bf#1} (#2) #3}

\def\EPL#1#2#3{{\sl Europhys. Lett.} {\bf#1} (#2) #3}
\def\NPB#1#2#3{{\sl Nucl. Phys.} {\bf B#1} (#2) #3}

\def\PRB#1#2#3{{\sl Phys. Rev.} {\bf B#1} (#2) #3}
\def\PLB#1#2#3{{\sl Phys. Lett.} {\bf #1B} (#2) #3}
\def\PLA#1#2#3{{\sl Phys. Lett.} {\bf #1A} (#2) #3}
\def\JMP#1#2#3{{\sl J. Math. Phys.} {\bf #1} (#2) #3}

\def\PTP#1#2#3{{\sl Prog. Theor. Phys.} {\bf #1} (#2) #3}

\def\ANYAS#1#2#3{{\sl Ann. New York Acad. Sci.} {\bf #1} (#2) #3}

\def\TMP#1#2#3{{\sl Theor. Mat. Phys.} {\bf #1} (#2) #3}
\def\JPA#1#2#3{{\sl J. Physics} {\bf A#1} (#2) #3}
\def\JPC#1#2#3{{\sl J. Physics} {\bf C#1} (#2) #3}
\def\JPCM#1#2#3{{\sl J. Physics Cond. Mat.} {\bf #1} (#2) #3}

\def\JSM#1#2#3{{\sl J. Soviet Math.} {\bf #1} (#2) #3}

\def\MPLB#1#2#3{{\sl Mod. Phys. Lett.} {\bf B#1} (#2) #3}
\def\JETP#1#2#3{{\sl Sov. Phys. JETP} {\bf #1} (#2) #3}

\def\ZPB#1#2#3{{\sl Z.Phys.} {\bf B#1} (#2) #3}
\def\sci#1#2#3{{\sl Science} {\bf #1} (#2) #3}

\def\SSC#1#2#3{{\sl Solid State Comm.} {\bf #1} (#2) #3}

\sectionnumstyle{arabic}
\footnoteskip=2pt
\footnotenumstyle{arabic}
\draft
\baselineskip=16pt
\def\scr{\scriptstyle}
\def\mp#1{{M^\prime_{#1}}}
\def\Th#1{{\theta({ #1 \over 2|U|})}}
\def\DTh#1{{4|U|\over 4U^2+ (#1)^2}}
\def\Re#1{{\rm Re}\left\lbrace#1\right\rbrace}
\def\inti{\int_{-\infty}^\infty}

\def\pl#1{(\putlab{#1})}
\def\da{\downarrow}
\def\up{\uparrow}
\def\ga{\alpha}
\def\gb{\beta}
\def\gc{\gamma}
\def\gd{\delta}
\def\ge{\epsilon}
\def\gs{\sigma}
\def\go{\omega}
\def\do{{d\go\over \go}}
\def\cd{c^\dagger}
\def\la{\lambda}
\def\La{\Lambda}
\def\tLa{{\tilde\Lambda}}
\def\tJ{{\tilde J}}
\def\p2{{\pi\over 2}}
\def\half{{1\over 2}}
\def\frac#1#2{{#1\over #2}}
\def\lh#1{\Lambda^h_{#1}}
\def\kh#1{{\rm sin}(k^h_{#1})}
\equation{H}{H = -\sum_{j=1}^L\sum_{\gs=\up ,\da}
\left(\cd_{j,\gs} c_{j+1,\gs} + \cd_{j+1,\gs} c_{j,\gs}\right)+
4U\sum_{j=1}^L (n_{j,\up}-{1\over 2})(n_{j,\da}-{1\over 2})\ .}

\equation{spin}{
S = \sum_{j=1}^L c^\dagger_{j,\up} c_{j,\da}\ , \quad
S^\dagger= \sum_{j=1}^L c^\dagger_{j,\da} c_{j,\up}\ ,\quad
S^z = \sum_{j=1}^L {1\over 2} (n_{j,\da}-n_{j,\up}) \ ,}

\equation{eta}{
\eta = \sum_{j=1}^L (-1)^jc_{j,\up} c_{j,\da}\ , \quad
\eta^\dagger = \sum_{j=1}^L (-1)^jc^\dagger_{j,\da} c^\dagger_{j,\up}\ ,
\quad \eta^z = {1\over 2}\sum_{j=1}^L (n_{j,\up}+n_{j,\da}  - 1)\ .}

\equation{eps}{\eqalign{
p_s(\la) &= \p2 - \int_{0}^\infty {d\go\over \go} {J_0(\go) {\rm
sin}(\go\la)\over {\rm cosh}(\go U)}\qquad 0<p_s(\la)<\pi\cr
\ge_s(\la) &= 2 \int_{0}^\infty {d\go\over \go} {J_1(\go) {\rm
cos}(\go\la)\over {\rm cosh}(\go U)}\ .\cr}}

\equation{epc}{\eqalign{
p_h(k^h) &= \p2-k^h - \int_{0}^\infty {d\go\over \go} {J_0(\go) {\rm
sin}(\go\ {\rm sin}(k^h))e^{-\go U}\over {\rm cosh}(\go U)}\cr
\ge_h(k^h) &= 2U+2\ {\rm cos}(k^h) + 2 \int_{0}^\infty {d\go\over \go}
{J_1(\go) {\rm cos}(\go\ {\rm sin}(k^h))e^{-\go U}\over {\rm cosh}(\go
U)}\ .\cr}}

\equation{antiholon}{\eqalign{
\ge_{ah}(k) &= \ge_h(k)\ ,\quad p_{ah}(k) = -\pi + p_h(k)\ .\cr}}

\equation{lambdastr}{
\La_{\alpha}^{m,j} = \La_{\alpha}^m - \frac{1}{2} (m+1-2j) u
\qquad \La_{\alpha}^m {\ \rm real}\qquad j=1,2, \ldots, m \ .}

\equation{pbc}{\eqalign{
& k_j L =
   2 \pi I_j
 + \sum_{n=1}^\infty \sum_{\alpha =1}^{M_n}
   \theta \left( \frac{\sin k_j - \Lambda_\alpha^n}{n |U|} \right)
 + \sum_{n=1}^\infty \sum_{\alpha =1}^{M^\prime_n}
   \theta \left( \frac{\sin k_j - \Lambda_\alpha^{\prime n}}
                      {n |U|} \right) \ ,\cr}}
\equation{pbc1b}{\sum_{j=1}^{N_e-2M^\prime}
   \theta \left( \frac{\Lambda_\alpha^n - \sin k_j}
                      {n |U|} \right)
 = 2 \pi  J_\alpha^n
   + \sum_{m=1}^\infty\sum_{\beta =1}^{M_m} \theta_{nm} \left(
     \frac{\Lambda_\alpha^n - \Lambda_\beta^m} {|U|} \right) \ ,}
\equation{pbc2}{\eqalign{
& L \left[ sin^{-1} \left( \Lambda_\alpha^{\prime n} + i n |U| \right)
  + sin^{-1} \left( \Lambda_\alpha^{\prime n} - i n |U|
\right) \right]\cr
& \hskip 1cm =
   2 \pi J_\alpha^{\prime n}
 + \sum_{j=1}^{N_e-2M^\prime}
   \theta \left( \frac{\Lambda_\alpha^{\prime n} - \sin k_j}
                      {n |U|} \right)
 + \sum_{m=1}^\infty\sum_{\beta =1}^{M_m^\prime} \Theta_{nm} \left(
     \frac{\Lambda_\alpha^{\prime n} - \Lambda_\beta^{\prime m}}
                      {|U|} \right) \ ,\cr}}

\equation{theta}{\theta_{nm}(x) = \cases{\theta ({x\over{|n-m|}}) +
2\ \theta ({x\over{|n-m|+2}})+\dots +2\ \theta ({x\over{n+m-2}}) +
\theta ({x\over{n+m}})&if $n\ne m$\cr\cr
2\ \theta ({x\over{2}}) + 2\ \theta ({x\over{4}})+\dots +
2\ \theta ({x\over{2n-2}})+\theta ({x\over{2n}})&if $n=m$\ \ .\cr}}

\equation{ineq}{\eqalign{
\mid J_\alpha^n \mid \leq & \half (N_e - 2 M^\prime
     - \sum_{m=1}^\infty t_{nm} M_m -1) \ ,\cr
\mid J_\alpha^{\prime n} \mid \leq & \half (L - N_e + 2 M^\prime
     - \sum_{m=1}^\infty t_{nm} M_m^\prime -1) \ ,\ -{L-1\over 2}\leq
I_j \leq {L-1\over 2}\ ,\cr}}

\equation{matrices}{I=\left(\matrix{1&0&0&0\cr 0&1&0&0\cr 0&0&1&0\cr
0&0&0&1\cr}\right)\ , \Pi=\left(\matrix{1&0&0&0\cr 0&0&1&0\cr 0&1&0&0\cr
0&0&0&1\cr}\right)\ .}

\equation{Sss}{S_{ss}(\kh{1}, \kh{2}) = {\Gamma({1-i\mu_1\over 2})
\Gamma(1+{i\mu_1\over 2}) \over \Gamma({1+i\mu_1\over 2})
\Gamma(1-{i\mu_1\over 2})}\ \left({\mu_1\over\mu_1-i} I - {i\over
\mu_1-i}\Pi\right)\ .}

\equation{parthol}{c_{j,\up}\leftrightarrow (-1)^j c^\dagger_{j,\up}\
,}

\equation{aeps}{\eqalign{
p_{cw}^p(\la) &= \pi - \int_{0}^\infty {d\go\over \go} {J_0(\go) {\rm
sin}(\go\la)\over {\rm cosh}(\go U)} = \pi + p_{cw}^h(\la)\ ,\cr
\ge_{cw}(\la) &= 2 \int_{0}^\infty {d\go\over \go} {J_1(\go) {\rm
cos}(\go\la)\over {\rm cosh}(\go U)} \ ,\cr}}

\equation{aepc}{\eqalign{
p_{sw}(k) &= k - \int_{0}^\infty {d\go\over \go} {J_0(\go) {\rm
sin}(\go\ {\rm sin}(k))e^{-\go |U|}\over {\rm cosh}(\go U)}\cr
\ge_{sw}(k) &= 2|U|-2\ {\rm cos}(k) + 2 \int_{0}^\infty {d\go\over \go}
{J_1(\go) {\rm cos}(\go\ {\rm sin}(k))e^{-\go |U|}\over {\rm cosh}(\go
U)}\ .\cr}}

\equation{liebwu}{\eqalign{
e^{i k_jL} &= \prod_{\alpha=1}^M {sin(k_j)-\Lambda_\alpha
+iU\over sin(k_j)-\Lambda_\alpha -iU}
\ \ , \ j=1,...,M+N\cr
\prod_{i=1}^{M+N} {sin(k_i)-\Lambda_\beta +iU\over
sin(k_i)-\Lambda_\beta -iU} &= \ - \prod_{\alpha=1}^M
{{\Lambda_\alpha- \Lambda_\beta +2iU} \over
{\Lambda_\alpha- \Lambda_\beta -2iU}} \ \ ,\ \ \beta = 1,...,M\ .\cr}}

\equation{lambdastr}{
\La_{\alpha}^{m,j} = \La_{\alpha}^m + i(m+1-2j)|U|
\qquad \La_{\alpha}^m {\ \rm real}\qquad j=1,2, \ldots, m \ .}

\equation{klstr1}{\eqalign{
k_\alpha^1 &= \sin^{-1}(\La_\alpha^{\prime \  m} +i m |U|)
\cr
k_\alpha^2 &= \sin^{-1}(\La_\alpha^{\prime \  m} + i(m-2) |U|)\ , \
k_\alpha^3 = \pi - k_\alpha^2\ ,\cr
k_\alpha^4 &= \sin^{-1}(\La_\alpha^{\prime \  m} +i (m-4) |U|)\ ,\
k_\alpha^5 = \pi - k_\alpha^4\ ,\cr
\ldots\cr
k_\alpha^{2m-2} &= \sin^{-1}(\La_\alpha^{\prime \  m} - i(m-2)|U|)\ ,\
k_\alpha^{2m-1} = \pi - k_\alpha^{2m-2}\cr
k_\alpha^{2m} &= \sin^{-1}(\La_\alpha^{\prime \  m} -im|U|)
\ .\cr}}

\equation{klstr2}{\La_{\alpha}^{\prime\ m,j} = \La_{\alpha}^{\prime\ m}
  +i(m+1-2j) |U| \ , \quad \La_{\alpha}^{\prime\ m}
{\ \rm real}\qquad j=1,2, \ldots, m \ . }

\equation{cons}{\
N_e = M_e + 2\  \sum_{m=1}^{\infty} m \  M^\prime_m \ ,
\hskip 6mm
M = \sum_{m=1}^{\infty} m (M_m + M_m^\prime) \ .}

\equation{GSEQ}{2L\ \Re{\arcsin(\La_\ga+ i |U|)}= 2\pi
J^{\prime 1}_\ga +\sum_{\gb =1}^{L\over 2}\Th{\La_\ga -
\La_\gb}\ , \ \ga =1,\ldots ,{L\over 2}.}

\equation{rho}{\rho_0(\La_\ga)= {1\over L(\La_{\ga +1} -
\La_{\ga})}\ .}

\equation{GSIE}{\eqalign{
\rho_0(\La) &= {1\over\pi}\Re{{1\over \sqrt{1-(\La +
i|U|)^2}}}  - {1\over 2\pi}\int_{-\infty}^\infty d\La^\prime\
{4|U|\over 4U^2+(\La-\La^\prime)^2}\ \rho_0(\La^\prime)\ ,\cr
&={1\over 2\pi}\inti d\go\ \exp(i\go\La){J_0(\go)\over 2\ {\rm
cosh}(U\go)}\ .\cr}}

\equation{E}{\eqalign{E &= -2\sum_{l=1}^{N_e-2M^\prime} \cos(k_l) -
4\sum_{n=1}^\infty\sum_{\ga=1}^{M^\prime_n} \Re{\sqrt{1-(\La^{\prime
n}_\ga+ i|U|n)^2}} -2 U N + U L\ ,\cr}}

\equation{P}{P = \sum_{j=1}^{N_e-2\mp{}}k_j +
\sum_{n=1}^\infty\sum_{\ga =1}^{\mp{n}}\left(2\Re{{\rm arcsin}(\La^{\prime
n}_\ga + i |U| n)} + \pi (n-1)\right)\ .}

\equation{GSE}{\eqalign{{E_{GS}\over L} &=
|U| - 4\inti d\La\ \Re{\sqrt{1-(\La + i|U|)^2}}\rho_0(\La)\cr
&= -|U|  - \inti \do {\exp(-|\go U|)\over {\rm
cosh}(\go U)} J_0(\go)J_1(\go)\ .\cr}}

\equation{BAECT}{2L\ \Re{\arcsin(\tLa_\ga+ i |U|)}= 2\pi
\tJ^{\prime 1}_\ga +\sum_{\gb =1}^{{L\over 2}+1} \Th{\tLa_\ga -
\tLa_\gb} - \sum_{j=1}^2 \Th{\tLa_\ga-\lh{j} }\ .}

\equation{diff1}{\eqalign{2L\ \Re{{1\over\sqrt{1-(\La_\ga+ i
|U|)^2}}}\left(\tLa_\ga - \La_\ga\right)&=
\pi + \Th{\tLa_\ga-\La_{min}}- \sum_{j=1}^2
\Th{\tLa_\ga-\lh{j} }\cr
&\hskip-40pt +\sum_{\gb =1}^{L\over 2} \DTh{\La_\ga -
\La_\gb}\left(\tLa_\ga - \La_\ga -(\tLa_\gb - \La_\gb)\right) \ .\cr}}

\equation{F}{F_{CT}(\La_\ga) = {\tLa_\ga - \La_\ga\over \La_{\ga
+1}-\La_\ga}\ , }

\equation{F0}{\eqalign{
&{F_{CT}(\La)\over \rho(\La)}\left(2\ \Re{{1\over \sqrt{1-(\La +
i|U|)^2}}}  - \int_{-\infty}^\infty d\La^\prime\
{4|U|\over 4U^2+(\La-\La^\prime)^2}\ \rho(\La^\prime)\right) \cr
&= 2\pi - \int_{-\infty}^\infty d\La^\prime\
{4|U|\over 4U^2+(\La-\La^\prime)^2}\ F_{CT}(\La^\prime) -
\sum_{j=1}^2 \Th{\La-\lh{j} }\ .\cr}}

\equation{F2}{F_{CT}(\La) = 1 - {1\over 2\pi}\inti d\La^\prime
\DTh{\La-\La^\prime} F_{CT}(\La^\prime) - {1\over 2\pi}\sum_{j=1}^2
\Th{\La-\lh{j} }\ .}

\equation{F3}{{\tilde F}(\go) = \pi\gd(\go) + {i\over \go}
{\exp(-i\go\lh{1})+ \exp(-i\go\lh{2}) \over 1+\exp(2|U\go|)\ .}}

\equation{fadtakh}{\gc(\mu)  = \inti {d\go\over i\go}{\exp(i\go\mu)
\over 1+\exp(|\go|)} = i\ln\left({\Gamma({1+i\mu\over 2})
\Gamma(1-{i\mu\over 2}) \over \Gamma({1-i\mu\over 2})
\Gamma(1+{i\mu\over 2})}\right)\ .}

\equation{F4}{F_{CT}(\La) = {1\over 2} - {1\over
2\pi}\left(\gamma\left({\La-\lh{1}\over 2|U|}\right) +
\gamma\left({\La-\lh{2}\over 2|U|}\right)\right)\ ,}

\equation{ECTold}{\eqalign{E_{CT} &= -4|U|
-4\sum_{\ga =1}^{{L\over 2}} \Re{
\sqrt{1-(\tLa_\ga + i|U|)^2}- \sqrt{1-(\La_\ga + i|U|)^2}}\cr
&\quad -4\Re{\sqrt{1-(\La_{min} + i|U|)^2}}
+4\sum_{j=1}^2\Re{\sqrt{1-(\lh{j} + i|U|)^2}}\cr
&= 4\inti\!\!\! d\La\ \Re{{\La + i|U|\over \sqrt{1-(\La + i|U|)^2}}}
F_{CT}(\La) +4\sum_{j=1}^2\left\lbrack\Re{\sqrt{1-(\lh{j} +
i|U|)^2}}-|U|\right\rbrack\cr
&= \ge_{cw}(\lh{1}) + \ge_{cw}(\lh{2})\ ,\cr}}

\equation{ECT}{\eqalign{E_{CT} &= 4\inti d\La\ \Re{{\La + i|U|\over
\sqrt{1-(\La + i|U|)^2}}} F_{CT}(\La)\cr
&\qquad +4\sum_{j=1}^2\left\lbrack\Re{\sqrt{1-(\lh{j} +
i|U|)^2}}-|U|\right\rbrack\cr
&= \ge_{cw}(\lh{1}) + \ge_{cw}(\lh{2})\ ,\cr}}

\equation{es}{\ge_{cw}(\la) = 2 \int_{0}^\infty {d\go\over \go}
{J_1(\go) {\rm cos}(\go\la)\over {\rm cosh}(\go U)}\ .}

\equation{PCT}{\eqalign{ P_{CT} &=
\inti d\La\ 2\Re{{1\over \sqrt{1-(\La + i|U|)^2}}} F_{CT}(\La)
-\sum_{j=1}^2\Re{{\rm arcsin}(\lh{j} + i|U|)}-\pi\cr
&= p^h_{cw}(\lh{1}) + p^h_{cw}(\lh{2})\ ,\cr}}

\equation{ps}{p_{cw}^h(\la) = - \int_{0}^\infty {d\go\over \go}
{J_0(\go) {\rm sin}(\go\la)\over {\rm cosh}(\go U)}\qquad
-\p2\leq p^h_{cw}(\la)\leq\p2\ .}

\equation{BAECS}{\eqalign{2L\ \Re{\arcsin(\tLa_\ga+ i |U|)}&= 2\pi
\tJ^{\prime 1}_\ga +\sum_{\gb =1}^{{L\over 2}} \Th{\tLa_\ga -
\tLa_\gb} - \sum_{j=1}^2 \Th{\tLa_\ga-\lh{j}
}\cr &+\qquad\theta_{12}({\tLa_\ga-\kappa\over |U|})\cr
2L\ \Re{\arcsin(\kappa+ 2i |U|)}&= \sum_{\gb =1}^{{L\over 2}}
\theta_{21}({\kappa - \tLa_\gb\over |U|}) - \sum_{j=1}^2
\theta_{21}({\kappa-\lh{j}\over |U|})\ .\cr}}

\equation{BAECS2}{\eqalign{
&F_{CS}(\La) = -{1\over 2\pi}\inti d\La^\prime\
{4|U|\over 4U^2+(\La-\La^\prime)^2}\ F_{CS}(\La^\prime)
+\theta_{12}({\La - \kappa\over |U|}) - \sum_{j=1}^2 \Th{\La-\lh{j}}
\cr & 2\ \Re{\arcsin(\kappa+ 2i |U|)}= \inti d\La\ \rho(\La)
\theta_{21}({\kappa - \La\over |U|}) - {1\over L}\sum_{j=1}^2
\theta_{21}({\kappa-\lh{j}\over |U|})\ .\cr}}

\equation{kappa}{\kappa = {\lh{1}+\lh{2}\over 2}\ .}

\equation{FCS}{F_{CS}(\La) = - {1\over
2\pi}\left(\gamma\left({\La-\lh{1}\over 2|U|}\right) +
\gamma\left({\La-\lh{2}\over 2|U|}\right)\right) +
{1\over 2\pi}\theta({\La-{\lh{1}+\lh{2}\over 2}\over |U|})\ ,}

\equation{EPCS}{\eqalign{E_{CS} &= \ge_{cw}(\lh{1}) +
\ge_{cw}(\lh{2})\ ,\cr P_{CS} &= p_{cw}^h(\lh{1}) + p_{cw}^p(\lh{2})\
,\cr}}

\equation{BAEes}{\eqalign{2L\ \Re{\arcsin(\tLa_\ga+ i |U|)}&= 2\pi
\tJ^{\prime 1}_\ga\! +\!\!\sum_{\gb =1}^{{L\over 2}} \Th{\tLa_\ga -
\tLa_\gb} - \Th{\tLa_\ga-\lh{}} + \theta({\tLa_\ga-\sin(k)\over |U|}),\cr
L\ k&= 2\pi {\tilde I}+ \sum_{\gb =1}^{{L\over 2}}
\theta({\sin(k) - \tLa_\gb\over |U|}) - \theta({\sin(k)-\lh{}\over
|U|})\ .\cr}}

\equation{Fes1}{
2\pi F_{\eta s}(\La) = -\inti d\La^\prime\
{4|U|\over 4U^2+(\La-\La^\prime)^2}\ F_{\eta s}(\La^\prime)
+\theta({\La - \sin(k)\over |U|}) - \Th{\La-\lh{}}\ ,}

\equation{Fes}{F_{\eta s}(\La) = {1\over\pi}
\arctan\left(\exp(\pi{\La-\sin(k)\over 2|U|})\right)-{1\over
4}-{1\over 2\pi}\gamma\left({\La -\lh{}\over 2|U|}\right)\ .}

\equation{EES}{\eqalign{E_{\eta s} &=
-2\cos(k) +4\inti d\La\ F_{\eta s}(\La) \Re{{\La+i|U|\over\sqrt{1-(\La
+ i|U|)^2} }}\cr
&\qquad +4\Re{\sqrt{1-(\lh{} + i|U|)^2}} -2|U|\cr
&= \ge_{cw}(\lh{}) + \ge_{sw}(k)\ ,\cr}}

\equation{ec}{\ge_{sw}(k) = 2|U|-2\cos(k)+ 2 \int_{0}^\infty {d\go\over
\go} {J_1(\go) {\rm cos}(\go\sin(k))\over {\rm cosh}(\go U)}
\exp(-|\go U|)\ .}

\equation{PES}{P_{\eta s}  = p_{cw}^h(\lh{}) + p_{sw}(k)\ ,}

\equation{pc}{p_{sw}(k) = k - \int_{0}^\infty {d\go\over \go}
{J_0(\go) {\rm sin}(\go\sin(k))\over {\rm cosh}(\go U)}\exp(-|\go U|)\ .}

\equation{BAESS}{\eqalign{2L\ \Re{\arcsin(\tLa_\ga+ i |U|)}&= 2\pi
\tJ^{\prime 1}_\ga\! +\sum_{\gb =1}^{{L\over 2}-1} \Th{\tLa_\ga -
\tLa_\gb} + \sum_{j=1}^2\theta({\tLa_\ga-\sin(k_j)\over |U|}),\cr
L\ k_j&= 2\pi {\tilde I_j}+ \sum_{\gb =1}^{{L\over 2}-1}
\theta({\sin(k_j) - \tLa_\gb\over |U|}) +
\theta({\sin(k_j)-\kappa\over |U|})\cr
\sum_{j=1}^2 \theta({\kappa -\sin(k_j)\over |U|})&=2\pi J^1_\ga\
.\cr}}

\equation{FSS}{\eqalign{
F_{SS}(\La) &= 1-{1\over 2\pi}\inti d\La^\prime\
{4|U|\over 4U^2+(\La-\La^\prime)^2}\ F_{SS}(\La^\prime)
+{1\over 2\pi}\sum_{j=1}^2\theta({\La - \sin(k_j)\over |U|})\cr
\Rightarrow F_{SS}(\La)&= {1\over\pi}
\sum_{j=1}^2\arctan\left(\exp(\pi{\La-\sin(k_j)\over
2|U|})\right)\cr .}}

\equation{EPSS}{\eqalign{E_{SS} &=\ge_{sw}(k_1) + \ge_{sw}(k_2)\ ,\cr
P_{SS} &=p_{sw}(k_1) + p_{sw}(k_2)\ ,\cr}}

\equation{BAEST}{\eqalign{2L\ \Re{\arcsin(\tLa_\ga+ i |U|)}&= 2\pi
\tJ^{\prime 1}_\ga\! +\sum_{\gb =1}^{{L\over 2}-1} \Th{\tLa_\ga -
\tLa_\gb} + \sum_{j=1}^2\theta({\tLa_\ga-\sin(k_j)\over |U|}),\cr
L\ k_j&= 2\pi {\tilde I_j}+ \sum_{\gb =1}^{{L\over 2}-1}
\theta({\sin(k_j) - \tLa_\gb\over |U|}) \ .\cr}}

\equation{Sgen}{S_{ab;cd}(\la_1,\la_2) =
S_1(\la_1,\la_2)\gd_{ac}\gd_{bd} +S_2(\la_1,\la_2) \gd_{ad}\gd_{bc}\
,}

\equation{ts}{S_t = S_1+S_2\ ,\ S_s = S_1-S_2\ .}

\equation{d12cs}{\eqalign{\gd_{CS}^{21}(\lh{1},\lh{2}) &=
-2L\ \Re{\arcsin(\lh{2}+ i |U|)}+\sum_{\gb =1}^{{L\over 2}}
\Th{\lh{2} - \tLa_\gb} - \Th{\lh{2}-\lh{1}}\cr
&\qquad + \theta_{12}({\lh{2}-\lh{1}\over 2|U|})\ .\cr}}

\equation{d0}{\gd_{0}(\lh{2}) =
-2L\ \Re{\arcsin(\lh{2}+ i |U|)}+\sum_{\gb =1}^{{L\over 2}}
\Th{\lh{2} - \La_\gb} \ , }

\equation{dcs}{\eqalign{\gd_{CS}(\lh{1},\lh{2})
&=\gd_{CS}^{21}(\lh{1},\lh{2}) - \gd_0(\lh{2})\cr
&=   -\inti d\La^\prime\ {4|U|\over 4U^2+(\La-\La^\prime)^2}\
F_{CS}(\La^\prime) +\theta_{12}({\lh{2} - \lh{1}\over 2|U|}) -
\Th{\lh{2}-\lh{1}}\cr
&= 2\pi\ F_{CS}(\lh{2}) = -\gamma(\mu_4)+\theta(\mu_4)\ ,\cr}}

\equation{d12ct}{\eqalign{\gd_{CT}^{21}(\lh{1},\lh{2}) &=
-2L\ \Re{\arcsin(\lh{2}+ i |U|)}+\sum_{\gb =1}^{{L\over 2}}
\Th{\lh{2} - \tLa_\gb} - \Th{\lh{2}-\lh{1}}\cr
&\quad +  \Th{\lh{2}-\La_{min}}+\pi\ .\cr}}

\equation{dct}{\eqalign{\gd_{CT}(\lh{1},\lh{2})
&=\gd_{CT}^{21}(\lh{1},\lh{2}) - \gd_0(\lh{2})\cr
&= 2\pi\ F_{CT}(\lh{2}) {\rm\ mod\ }2\pi\ = \pi-\gamma(\mu_4){\rm\ mod\
}2\pi\ ,\ \mu_4 = {\lh{2}-\lh{1}\over 2|U|}>0\ .\cr}}

\equation{scc}{\eqalign{S_s &= \exp(i\gd_{CS})=-{\mu_4 -i\over \mu_4 +i}\
{\Gamma({1+i\mu_4\over 2})\Gamma(1-{i\mu_4\over 2}) \over \Gamma({1-i\mu_4\over
2})
\Gamma(1+{i\mu_4\over 2})}\cr
S_t &= -{\Gamma({1+i\mu_4\over 2})\Gamma(1-{i\mu_4\over 2}) \over
\Gamma({1-i\mu_4\over 2}) \Gamma(1+{i\mu_4\over 2})}\ .\cr}}

\equation{Scc}{S_{cc}(\lh{1},\lh{2}) = -{\Gamma({1+i\mu_4\over 2})
\Gamma(1-{i\mu_4\over 2}) \over \Gamma({1-i\mu_4\over 2})
\Gamma(1+{i\mu_4\over 2})}\ \left({\mu_4\over\mu_4+i} I + {i\over
\mu_4+i}\Pi\right)\ .\ }

\equation{d12es}{\eqalign{\gd_{\eta s}^{21}(\lh{},k) &=
-2L\ \Re{\arcsin(\lh{}+ i |U|)}+\sum_{\gb =1}^{{L\over 2}}
\Th{\lh{} - \tLa_\gb} +  \theta({\lh{}-\sin(k)\over |U|}) .\cr}}

\equation{des}{\eqalign{\gd_{\eta s}(\lh{},k)
&=\gd_{\eta s}^{21}(\lh{},k) - \gd_0(\lh{}) =  2\pi\ F_{\eta
s}(\lh{2}) {\rm\ mod\ }2\pi\cr
&= 2\ \arctan(\exp(\pi\mu_3))-{\pi\over 2}\ ,\
\mu_3={\lh{}-\sin(k)\over 2|U|}>0\ .\cr}}

\equation{Ses}{S_{\eta s}(\lh{},k) = -i\ {1+i\ \exp({\pi}\mu_3)\over 1-i\
\exp({\pi}\mu_3)} I\ .}

\equation{d12se}{\eqalign{\gd_{s\eta}^{21}(\lh{},k) &=
Lk-\sum_{\gb =1}^{{L\over 2}}\theta({\sin(k) - \tLa_\gb\over |U|}) +
\theta({\sin(k) - \lh{}\over |U|})\ .\cr}}

\equation{dse}{\eqalign{\gd_{s\eta }(\lh{},k)
&=\gd_{s\eta }^{21}(\lh{},k) - {\tilde\gd}_0(k)\cr
&= \inti d\La^\prime\ {2|U|\over U^2+(sin(k)-\La^\prime)^2}F_{\eta
s}(\La^\prime) + \theta({\sin(k)-\lh{}\over |U|})\cr
&=2\ \arctan(\exp(\pi\mu_2))-{\pi\over 2}\ ,\ \mu_2={\sin(k)-\lh{}\over
2|U|}>0\ .\cr}}

\equation{d12st}{\gd^{21}_{ST}(k_1,k_2) = Lk_2 - \sum_{\ga=1}^{{L\over
2}-1} \theta({\sin(k_2) - \tLa_\ga\over |U|})\ .}

\equation{dstLONG}{\eqalign{\gd_{ST}(k_1,k_2)&=\gd_{ST}^{21}(k_1,k_2) -
\gd_0(k_2)\cr
&= \inti d\La\ {2|U|\ober U^2+(\sin(k_2)-\La)^2}F_{ST}(\La) +
\theta({\sin(k_2) - \La_{min}\over |U|})\cr
&= \gamma(\mu)\ {\rm mod}\ 2\pi\ ,\ \mu = {\sin(k_2)-\sin(k_1)\over
2|U|}>0\ .\cr}}

\equation{dst}{\eqalign{\gd_{ST}(k_1,k_2)&= \gamma(\mu_1)\ {\rm mod}\
2\pi\ ,\ \mu_1 = {\sin(k_2)-\sin(k_1)\over 2|U|}>0\ .\cr}}

\equation{dss}{\eqalign{\gd_{SS}(k_1,k_2)&= \gamma(\mu_1)-2\
\arctan(\mu_1)+\pi\ {\rm mod}\ 2\pi\ ,\cr}}

\equation{SM}{S=\left(\matrix{S_{ss}(\mu_1)&0&0&0\cr
0&S_{s\eta}(\mu_2)&0&0\cr 0&0&S_{\eta s}(\mu_3)&0\cr
0&0&0&S_{cc}(\mu_4)\cr}\right)\ .}


\equation{GSREP}{\eqalign{L\ k_l &= 2\pi\ I_l - \sum_{\ga=1}^{L\over
2} \theta({\sin(k_l)-\La_\ga\over U})\cr
\sum_{l=1}^L \theta({\La_\ga - \sin(k_l) \over U})&=
2\pi\ J^1_\ga - \sum_{\gb=1}^{L\over
2} \theta({\La_\ga-\La_\gb\over 2U})\ ,\cr}}

\equation{GSREP2}{\eqalign{\rho(k) &= {1\over 2\pi} + {1\over
2\pi}\cos(k) \inti d\La {2U\over U^2 + (\sin(k)-\La)^2} \gs(\La)\cr
\gs(\La)&= {1\over 2\pi} \int_{-\pi}^\pi dk {2U\over U^2 +
(\sin(k)-\La)^2} \rho(k) -{1\over 2\pi}\inti d\La^\prime
{4U\over 4U^2 + (\La-\La^\prime)^2} \gs(\La^\prime)\ .\cr}}

\equation{rBAECT}{\eqalign{
L{\tilde k}_l &= 2\pi {\tilde I}_l- \sum_{\ga=1}^{L-2\over
2} \theta({\sin({\tilde k}_l)-\tLa_\ga\over U})\cr
\sum_{l=1}^L \theta({\tLa_\ga - \sin({\tilde k}_l) \over U})
-\sum_{j=1}^2 \theta({\tLa_\ga - \kh{j} \over U})&=
2\pi\ {\tilde J}_\ga - \sum_{\gb=1}^{L-2\over
2} \theta({\tLa_\ga-\tLa_\gb\over 2U})\ .\cr}}

\equation{rFCT}{\eqalign{F^c_{CT}(k) &={1\over 2\pi}\inti d\La
{2U\over U^2+(\sin(k) -\La)^2} F_{CT}^s(\La)\cr
F^s_{CT}(\La) &=-1 -{1\over 2\pi}\inti d\La^\prime
{4U\over 4U^2+(\sin(k) -\La^\prime)^2} F_{CT}^s(\La^\prime) + {1\over
2\pi}\sum_{j=1}^2 \theta({\La - \kh{j} \over U})\ .\cr}}

\equation{rFCT2}{\eqalign{F^s_{CT}(\La) &= -1 + {1\over
\pi}\sum_{j=1}^2 \arctan(\exp(\pi{\La-\kh{j}\over 2U}))\cr
F^c_{CT}(k) &= -{1\over 2} + {1\over 2\pi} \sum_{j=1}^2 \gc\left(
\sin(k)-\kh{j}\over 2U\right)\ ,\cr}}

\equation{rEPCT}{\eqalign{E_{CT}(k^h_1,k^h_2) &= 4U +
2\sum_{j=1}^2\cos(k^h_j) +  2\int_{-\pi}^\pi dk\ \sin(k)
F_{CT}^c(k)=\ge_h(k^h_1) + \ge_h(k^h_2)\ ,\cr
P_{CT}(k^h_1,k^h_2) &= -\sum_{j=1}^2 k^h_{j} + \int_{-\pi}^\pi dk\
F_{CT}^c(k) = p_h(k^h_1) + p_h(k^h_2)\ ,\cr}}

\equation{repc}{\eqalign{
p_h(k) &= \p2-k - \int_{0}^\infty {d\go\over \go} {J_0(\go) {\rm
sin}(\go\ {\rm sin}(k))e^{-\go |U|}\over {\rm cosh}(\go U)}\cr
\ge_h(k) &= 2U+2\ {\rm cos}(k) + 2 \int_{0}^\infty {d\go\over \go}
{J_1(\go) {\rm cos}(\go\ {\rm sin}(k))e^{-\go |U|}\over {\rm cosh}(\go
U)}\ .\cr}}

\equation{rFCS}{\eqalign{F^s_{CS}(\La) &= -1 + {1\over
\pi}\sum_{j=1}^2 \arctan(\exp(\pi{\La-\kh{j}\over 2U}))\cr
F^c_{CS}(k) &= -{1\over \pi}\arctan({\sin(k)-{\kh{1}+\kh{2}\over
2}\over U}) + {1\over 2\pi} \sum_{j=1}^2 \gc\left(
\sin(k)-\kh{j}\over 2U\right)\ ,\cr}}

\equation{rPCT}{\eqalign{E_{CS}(k^h_1,k^h_2) &=
\ge_{h}(k^h_1) + \ge_{ah}(k^h_2)\ ,\cr
P_{CS}(k^h_1,k^h_2) &= p_{h}(k^h_1) + p_{ah}(k^h_2)\ ,\cr}}

\equation{rFes}{\eqalign{F^s_{\eta s}(\La) &= {1\over
\pi}\arctan(\exp(\pi{\La-\kh{}\over 2U}))-{1\over 4}
-{1\over 2\pi} \gc\left(\La-\lh{}\over 2U\right)\ ,\cr
F^c_{\eta s}(k) &= {1\over 4}+{1\over
\pi}\arctan(\exp(\pi{\sin(k)-\lh{}\over 2U})) + {1\over 2\pi}
\gc\left(\sin(k)-\kh{}\over 2U\right)\ .\cr}}

\equation{rGSE}{\eqalign{{E_{GS}(U>0)\over L} &=
 -U - \inti \do {\exp(-|\go| U)\over {\rm
cosh}(\go U)} J_0(\go)J_1(\go)\ .\cr}}

\equation{rEetas}{\eqalign{E_{\eta s}(k^h,\lh{})
&=\ge_h(k^h) + \ge_s(\lh{})\ ,\cr
P_{\eta s}(k^h,\lh{}) &= p_h(k^h) + p_s(\lh{})\ ,\cr}}

\equation{reps}{\eqalign{
p_s(\la) &= \p2 - \int_{0}^\infty {d\go\over \go} {J_0(\go) {\rm
sin}(\go\la)\over {\rm cosh}(\go U)} \ ,\cr
\ge_s(\la) &= 2 \int_{0}^\infty {d\go\over \go} {J_1(\go) {\rm
cos}(\go\la)\over {\rm cosh}(\go U)} \ .\cr}}

\equation{rE}{\eqalign{E &= -2\sum_{l=1}^{N_e-2M^\prime} \cos(k_l) +
4\sum_{n=1}^\infty\sum_{\ga=1}^{M^\prime_n} \Re{\sqrt{1-(\La^{\prime
n}_\ga+ i|U|n)^2}}-2 U N_e + U L\ ,\cr}}

\equation{rP}{P = \sum_{j=1}^{N_e-2\mp{}}k_j -
\sum_{n=1}^\infty\sum_{\ga =1}^{\mp{n}}\left(2\Re{{\rm arcsin}(\La^{\prime
n}_\ga + i |U| n)} + \pi (n-1)\right)\ .}

\equation{rFST}{\eqalign{
F^s_{ST}(\La) &= {1\over 2} - {1\over 2\pi} \sum_{j=1}^2 \gc\left(
\La-\lh{j}\over 2U\right)\ ,\cr
F^c_{ST}(k) &= {1\over \pi}\sum_{j=1}^2
\arctan(\exp(\pi{\sin(k)-\lh{j}\over 2U})) . \cr}}

\equation{rFSS}{\eqalign{ F^s_{SS}(\La) &= -{1\over 2\pi} \sum_{j=1}^2
\gc\left( \La-\lh{j}\over 2U\right)+ {1\over
\pi}\arctan({\La-{\lh{1}+\lh{2}\over 2}\over U}) \ ,\cr
F^c_{SS}(k) &\equiv F^c_{ST}(k) . \cr}}

\equation{rEPspin}{\eqalign{E(\lh{1},\lh{2})
&=\ge_s(\lh{1}) + \ge_s(\lh{2})\ ,\cr
P(\lh{1},\lh{2}) &= p_s(\lh{1}) + p_s(\lh{2})\ ,\cr}}

\equation{rphase1}{\eqalign{
\gd_{CT} &= 2\pi\ F_{CT}^c(k^h_2) +\pi=\ i\
\ln\left({\Gamma({1+i\mu_4\over 2}) \Gamma(1-{i\mu_4\over 2}) \over
\Gamma({1-i\mu_4\over 2}) \Gamma(1+{i\mu_4\over 2})}\right),\
\mu_4={\kh{2}-\kh{1}\over 2U}>0\ ,\cr
\gd_{CS} &= 2\pi\ F_{CS}^c(k^h_2) +\pi=\pi-2\ \arctan(\mu_4) + i\
\ln\left({\Gamma({1+i\mu_4\over 2})
\Gamma(1-{i\mu_4\over 2}) \over \Gamma({1-i\mu_4\over 2})
\Gamma(1+{i\mu_4\over 2})}\right)\ ,\cr}}

\equation{rphase2}{\eqalign{
\gd_{ST} &= 2\pi\ F_{ST}^s(\lh{2}) =\pi- i\
\ln\left({\Gamma({1+i\mu_1\over 2}) \Gamma(1-{i\mu_1\over 2}) \over
\Gamma({1-i\mu_1\over 2}) \Gamma(1+{i\mu_1\over 2})}\right),\
\mu_1={\lh{2}-\lh{1}\over 2U}>0\ ,\cr
\gd_{SS} &= 2\pi\ F_{SS}^s(\lh{2}) =2\ \arctan(\mu_1)- i\
\ln\left({\Gamma({1+i\mu_1\over 2}) \Gamma(1-{i\mu_1\over 2}) \over
\Gamma({1-i\mu_1\over 2}) \Gamma(1+{i\mu_1\over 2})}\right)\ ,\cr}}

\equation{rphase3}{\eqalign{
\gd_{\eta s} &= 2\pi\ F_{\eta s}^c(k^h)+\pi =2\
\arctan(\exp(\pi\mu_3))- \p2\ , \mu_3 = {\sin(k^h)-\lh{}\over 2U}>0\
,\cr \gd_{s\eta } &= 2\pi\ F_{\eta s}^s(\lh{}) =2\
\arctan(\exp(\pi\mu_2))- \p2\ , \mu_2 = {\lh{}-\sin(k^h)\over 2U}>0\
.\cr}}

\equation{rScc}{S_{cc}(k^h_1,k^h_2) = {\Gamma({1-i\mu_4\over 2})
\Gamma(1+{i\mu_4\over 2}) \over \Gamma({1+i\mu_4\over 2})
\Gamma(1-{i\mu_4\over 2})}\ \left({\mu_4\over\mu_4-i} I - {i\over
\mu_4-i}\Pi\right)\ .}

\equation{rSss}{S_{ss}(\lh{1}, \lh{2}) = -{\Gamma({1+i\mu_1\over 2})
\Gamma(1-{i\mu_1\over 2}) \over \Gamma({1-i\mu_1\over 2})
\Gamma(1+{i\mu_1\over 2})}\ \left({\mu_1\over\mu_1+i} I + {i\over
\mu_1+i}\Pi\right)\ ,}

\equation{rSse}{S_{s\eta}(\lh{},k) = -i\ {1+i\ \exp({\pi}\mu_2)\over 1-i\
\exp({\pi}\mu_2)} I\ ,}

\equation{rSes}{S_{\eta s}(\lh{},k) = -i\ {1+i\ \exp({\pi}\mu_3)\over 1-i\
\exp({\pi}\mu_3)} I\ ,}

\equation{holes}{\eqalign{
&\rho(k) + {\bar \rho}(k) = {1\over 2\pi} - {\cos(k)\over 2\pi}
\sum_{n=1}^\infty \int_{-\infty}^\infty d\Lambda\ {2n|U|\over
n^2U^2+(\sin k - \Lambda)^2}(\gs_n(\La) + \gs^\prime_n(\La))\cr
& {\bar\gs}_n(\La) = {1\over 2\pi}\int_{-\pi}^\pi dk\ {2n|U|\over
n^2U^2 + (\Lambda - \sin k)^2} \rho(k) - \sum_{m=1}^\infty
A_{nm}*\gs_m\bigg|_\La\cr
&{\bar\gs}^\prime_n(\La) = {1\over\pi}\Re{\!\!
1\over\sqrt{1-(\La+in|U|)^2}\!} - \int_{-\pi}^\pi\!\!\!\! {dk\over
2\pi} {2n|U|\rho(k)\over n^2U^2+(\sin(k)-\La)^2} - \sum_{m=1}^\infty
A_{nm}*\gs^\prime_m\bigg|_\La ,\cr}}

\equation{TBA}{\eqalign{
\ln(\zeta(k)) &= {-2\ \cos(k)-\mu\over T} + \sum_{n=1}^\infty{1\over
2\pi} \inti d\la {2n|U|\over n^2U^2 + (\sin(k)-\La)^2}\ln({1+{1\over
\eta^\prime_n}\over1+{1\over \eta_n}})\ ,\cr
\ln(1+\eta_n(\La)) &= \sum_{n=1}^\infty
A_{nm}*\ln(1+{1\over\eta_m})\bigg|_\La\cr
&\qquad  +{1\over 2\pi}\int_{-\pi}^\pi
dk {2n|U|\cos(k)\over n^2U^2+(\sin(k)-\La)^2}\ln(1+{1\over\zeta(k)})\ ,\cr
\ln(1+\eta^\prime_n(\La)) &= {-4\Re{\sqrt{1-(\La+in|U|)^2}}-2n\mu\over T}
+\sum_{n=1}^\infty A_{nm}*\ln(1+{1\over\eta^\prime_m})\bigg|_\La\cr
&\qquad\qquad +{1\over 2\pi}\int_{-\pi}^\pi dk {2n|U|\cos(k)\over
n^2U^2+(\sin(k)-\La)^2}\ln(1+{1\over\zeta(k)})\ .\cr }}

\equation{dresseden}{\eqalign{
\ge_1^\prime(\La) &= -2 \int_{0}^\infty {d\go\over \go} {J_1(\go) {\rm
cos}(\go\La)\over {\rm cosh}(\go U)} , \cr
\kappa(k) &= 2|U|-2\ {\rm cos}(k) + 2 \int_{0}^\infty {d\go\over \go}
{J_1(\go) {\rm cos}(\go\ {\rm sin}(k))e^{-\go |U|}\over {\rm cosh}(\go
U)}\ ,\cr
\ge_n^\prime (\La) &= 0\ \forall\ n\geq 2\ ,\quad \ge_n(\La) = 0\
\forall\ n\geq 1\ .\cr}}

\pagenumstyle{blank}
\footnoteskip=2pt
\line{\it August 1993\hfil ITP-SB-93-45}
\vskip4em
\baselineskip=32pt
\begin{center}
{\bigsize{\sc $SU(2)\times SU(2)$-invariant Scattering Matrix\\
of the Hubbard Model}}
\end{center}
\vfil
\baselineskip=16pt

\begin{center}
{\bigsize
Fabian H.L.E\sharps ler\footnote[$\ \flat$]{\sc e-mail:
fabman@avzw01.physik.uni-bonn.de}}\vskip 0.5cm
{\it Physikalisches Institut der Universit\"at Bonn\vskip 4pt
Nussallee 12, 53115 Bonn}
{\bigsize
\vskip .5cm
and \vskip .5cm
Vladimir E. Korepin\footnote[$\ \sharp$]{\sc e-mail:
korepin@max.physics.sunysb.edu}\vskip .5cm}

\it Institute for Theoretical Physics\vskip 4pt
State University of New York at Stony Brook\vskip 4pt
Stony Brook, NY~~11794-3840
\end{center}

\vfil

\centertext{\bfs \bigsize ABSTRACT}
\vskip\belowsectionskip

\begin{narrow}[4em]
\noindent
We consider the one-dimensional half-filled Hubbard model.
We show that the excitation spectrum is given by the scattering states
of four elementary excitations, which form the fundamental
representation of $SU(2)\times SU(2)$.
We determine the exact two-particle Scattering matrix, which
a solution of the Yang-Baxter equation and reflects the $SO(4)$
symmetry of the model. The results for repulsive and attractive
Hubbard model are related by an interchange of spin and charge degrees
of freedom.
\end{narrow}
\vfil

\break

\pagenumstyle{arabic}

{\sc\section{Introduction}}
Strongly correlated electronic systems are currently under intense
study in relation with the phenomenon of high-$T_c$ superconductivity.
The two-dimensional Hubbard model is believed to be the most promising
candidate for an electronic theory of superconductivity. It is
believed to share important features with its one-dimensional
analog\upref pwa/. An important problem in understanding the dynamics
of the Hubbard model is the separation of spin and charge in one\upref
woynar1, woynar2/ and two dimensions\upref pwa2/. A particularly
interesting feature of the Hubbard model is that in one dimension it
can be solved exactly by means of the Bethe Ansatz\upref lieb/. Thus
it is possible to obtain a complete and unambiguous picture of the
dynamics in one dimension. The main purpose of this paper is to
use the Bethe Ansatz solution to demonstrate that the one-dimensional
Hubbard model is a factorizable scattering theory of four
quasiparticles. These elementary excitations transform in the
fundamental representation of $SU(2)\times SU(2)$.

The Hubbard hamiltonian is given by the following expression

$$\putequation{H}$$

Here $\cd_{j,\gs}$ are canonical fermionic creation operators on the
lattice, $j$ labels the sites of a chain of even length $L$, $\gs$
labels the spin degrees of freedom, $U$ is the coupling constant, and
$n_{j,\gs} = \cd_{j,\gs}c_{j,\gs}$ is the number operator for spin
$\gs$ on site $j$. The hamiltonian \pl{H} is invariant under a
SO(4)$=$SU(2)$\times$SU(2)/$Z_2$ algebra\upref hl,yang1,yang2, per,
aff/. The two SU(2) algebras are the spin SU(2) generated by
$$\putequation{spin}$$
and the $\eta$-SU(2) algebra
$$\putequation{eta}$$
\noindent All $6$ generators commute with the hamiltonian \pl{H}.
Clearly the operator $S^z+\eta^z$ has only integer eigenvalues (as $L$
is even) and all half-odd integer representations of $SU(2)\times
SU(2)$ are projected out. Thus the symmetry algebra is $SO(4)$ as
asserted above.\par
The nested Bethe Ansatz for the Hubbard model\upref lieb/ provides
eigenstates of the hamiltonian that are parametrized by sets of
so-called spectral parameters $k_j$ and $\La_\ga$. These parameters
are subject to the Lieb-Wu equations
$$\putequation{liebwu}$$
These equations are the basis for the determination of ground state
and excitation spectrum in the thermodynamic limit.\\
There has been substantial previous work on the excitation spectrum of
the repulsive, half-filled Hubbard model. A.A. Ovchinnikov calculated
the spin-triplet excitation\upref ov/, which was then re-examined by
T.C. Choy and W. Young in [\putref{cy}]. In his series of excellent
papers\upref woynar1,woynar2,woynar3/ F. Woynarovich gave a very
detailed analysis of both spin and charge excitations. The attractive
case was studied by B. Sutherland, who investigated the Bethe Ansatz
structure of the ground state, and by F. Woynarovich, who determined
ground state and excitation spectrum from the known results for the
repulsive case by using discrete symmetries of the hamiltonian\upref
woynar/. This method, while providing a nice picture of the relations
between the Bethe Ans\"atze for repulsion and attraction has the
drawback of making assumptions about transformation properties of
spectral parameters under the map connecting the $U>0$ and $U<0$
cases. In section $2$ we present a direct Bethe Ansatz analysis of the
ground state and excitation spectrum in the two-particle sector of the
attractive Hubbard model. Our direct computation as expected confirms
Woynarovich's results for the dispersion relations\upref woynar/. The
main purpose of this section is to determine the so-called
``shift-functions''\upref vladb,bik2/, which are needed to evaluate the
two-particle S-matrix. In section $3$ we give a quasiparticle
interpretation of the excitation spectrum. All excited states are {\sl
scattering states} of only four quasiparticles (or elementary
excitations), which form the fundamental representation of
$SU(2)\times SU(2)$. Two of these elementary excitations carry spin
but no charge, and two carry charge but no spin. This reflects the
separation of spin and charge degrees of freedom in the
one-dimensional Hubbard model. In section $4$ we evaluate the exact
S-matrix of the attractive Hubbard model. It is a solution of the
Yang-Baxter equation, which implies the two-particle reducibility of
the $N$-body S-matrix. In section $5$ we give the
results for the repulsive case, and in section $6$ we show that
attractive and repulsive Hubbard model are related by an
interchange of spin-and charge degrees of freedom. This relation holds
both on the level of the quasiparticle spectrum and on the level of
the S-matrix.
Finally section $7$ is devoted to a summary and discussion of our
results. The appendix deals with the problem of showing that all
excitations in the $N$-particle sector can be interpreted as
scattering states of the four elementary excitations.
\vskip .4cm

{\sc\section{Attractive Hubbard Model}}
In this section we construct the ground state and all excitations in
the two-particle sector for the attractive case $U<0$.
Starting point are the Lieb-Wu equations \pl{liebwu}.
The solutions of \pl{liebwu} can be split into three different types
of subsets (strings), which are\upref taka/
\item{ (1)}
  a single real momentum $k_i$
\item{ (2)}
  $m$ $\La_\alpha$'s combine into a string-type configuration
  (`$\La$-strings'); this includes the case $m=1$, which is just a
  single real $\La_\alpha$
\item{(3) }
  $2m$ $k_i$'s and $m$ $\La_\alpha$'s combine into a different
   string-type configuration (`$k$-$\La$-strings')

For a $\La$-string of length $m$ the rapidities involved are

$$\putequation{lambdastr}$$

The $k$'s and $\La$'s involved in a $k$-$\La$-string are

$$\putequation{klstr1}$$

and

$$\putequation{klstr2}$$

\noindent
Equations \pl{lambdastr} - \pl{klstr2} are valid up to exponential
corrections of order ${\cal O}(\exp(-\delta L))$, where
$\delta$ is real and positive (as long as the real parts of the
spectral parameters are much smaller than $L$). Note that the
$k$-$\La$-strings for the attractive Hubbard model are different
from the repulsive case treated by M. Takahashi in [\putref{taka}].

We now consider a solution of \pl{liebwu} with $M_n$ strings of type
\pl{lambdastr} of length $n$, $M_n^\prime$ $k$-$\Lambda$-strings
\pl{klstr1}, \pl{klstr2} of length $n$, and a total number of
electrons $N_e$. Inserting the prescription \pl{lambdastr} -
\pl{klstr2} into the Lieb-Wu equations \pl{liebwu} and taking the
logarithm we arrive at

$$\putequation{pbc}$$
$$\putequation{pbc1b}$$
$$\putequation{pbc2}$$

where $L=2\times {\rm odd}$ is the even length of the lattice, $I_j$,
$J^n_\ga$, and $J^{\prime n}_\ga$ are integer or half-odd integer
numbers, $M^\prime=\sum_{n=1}^\infty nM^\prime_n$, $\theta(x)=2{\ \rm
arctan}(x)$, and
$$\putequation{theta}$$
The integer (half-odd integer) numbers in \pl{pbc}-\pl{pbc2} have the
ranges
$$\putequation{ineq}$$
where $t_{mn} = 2{\rm min}\{m,n\}-\gd_{mn}$. Each set of ``integers''
$\{I_j\},\ \{J^n_\ga\},\ \{J^{\prime n}_\ga\}$ is in one-to-one
correspondence with a set of spectral parameters, which in turn
unambiguously specifies one eigenstate of the hamiltonian \pl{H}.
Energy and momentum of a solution of the system \pl{pbc}-\pl{pbc2} are
found to be
$$\putequation{E}$$
$$\putequation{P}$$
Equations \pl{pbc} - \pl{P} can now be used to determine the ground
state energy and to construct excitations. The ground state can be found
by taking the zero temperature limit of the Thermodynamic Bethe Ansatz
based on \pl{pbc}-\pl{E}, which can be constructed analogously to the
repulsive case\upref taka/ (see the Appendix of this paper).
An important element in our construction of excitations over the
ground state is a theorem proved in [\putref{eks}], which states that
\item{(i)} All eigenstates of the hamiltonian \pl{H} with finite
spectral parameters, $N_\da < N_\up$, and $N_e\leq L$ are {\sl lowest
weight} states of the $SO(4)$ algebra \pl{spin},\pl{eta}. These states
are called {\sl regular}  Bethe states.
\item{(ii)} The set of states obtained by acting with the $SO(4)$
raising-operators on the regular Bethe states is complete and forms a
basis of the electronic Hilbert space.\\ \noindent
This theorem implies that excitations constructed by means of the
Bethe Ansatz are all lowest weight states of the $SO(4)$ algebra.
Furthermore all $SO(4)$-descendants of these excitations must be taken
into account in order to obtain a complete set of excited states.
\vskip .4cm
{\sl\centerline{Ground State}}
\vskip .4cm
The ground state for the half-filled band is characterized
by choosing $M^{\prime}_1 = {L\over 2}$, $M_n = 0\ \forall n$,
$M^{\prime }_n = 0\ \forall n\geq 2$, and filling all $L\over 2$
vacancies given by \pl{ineq} for the integers $J^{\prime 1}_\ga$
symmetrically between $-{L-2\over 4}$ and ${L-2\over 4}$.
The Lieb-Wu equations \pl{pbc}-\pl{pbc2} for the ground state are
$$\putequation{GSEQ}$$
Subtracting equations \pl{GSEQ} for consecutive $\ga$'s and using
$J^{\prime1}_{\ga+1} - J^{\prime1}_{\ga} = 1$ we obtain an equation for
the finite-interval density
$$\putequation{rho}$$
\noindent In the thermodynamic limit this equation turns into an
integral equation for the ground state density $\rho_0(\La)$ (which is
defined as the limit of \pl{rho})
$$\putequation{GSIE}$$
The ground state energy per site can be evaluated by using \pl{E}
$$\putequation{GSE}$$
Next we will determine excitations over the ground state \pl{GSIE}
by means of the so-called ``shift-function'' method, which is
explained in detail in [\putref{bik2}] and [\putref{vladb}].
\vfill\newpage
{\sl\centerline{Charge-Triplet Excitation}}
\vskip .4cm
The lowest weight state of the charge triplet ($\eta^z = -1$) is
is obtained by choosing $M^{\prime }_1 = {L\over 2}-1$. The allowed
range \pl{ineq} of the half-odd integers $J^{\prime 1}_\ga$ is
$[-{L\over 4}, {L\over 4}]$, so that there are ${L\over 2}+1$
vacancies and thus two ``holes''. We denote the spectral parameters
corresponding to these holes by $\lh{1}$ and $\lh{2}$. The Lieb-Wu
equations read
$$\putequation{BAECT}$$
\noindent Here the tilde indicates that the spectral parameters
and the integers have changed as compared to their ground state
distributions. Our convention is $\tJ^\prime_\ga -
J^\prime_\ga={1\over 2}$ for $\ga=1\ldots {L\over 2}$
\footnote{The results for energy, momentum, and scattering phase are
of course independent of our choice of convention.}. As compared to
the ground state there is one more vacancy in the excited state and
thus one more $\tJ^{\prime}_\ga$, which in our convention is the {\sl
minimal} allowed half-odd integer $-{L\over 4}$. We denote the
corresponding spectral parameter by $\La_{min}$ and note that in the
thermodynamic limit $\La_{min}\rightarrow -\infty$.
Subtracting \pl{BAECT} from \pl{GSEQ} and using that $\tLa_\ga -
\La_\ga = {\cal O}({1\over L})$ we find

$$\putequation{diff1}$$

We now define the ``shift'' function $F_{CT}$ for the charge-triplet
according to

$$\putequation{F}$$

and take the thermodynamic limit of \pl{diff1}

$$\putequation{F0}$$

To order ${\cal O}(L^{-1})$ $\rho(\La)$ is the same as $\rho_0(\La)$,
so that we can use \pl{GSIE} in \pl{F0}, obtaining an integral
equation for $F_{CT}$

$$\putequation{F2}$$

Equation \pl{F2} has the solution

$$\putequation{F4}$$
\noindent where $\gamma(\mu)$ is defined as\upref natan1/

$$\putequation{fadtakh}$$

The excitation energy follows from \pl{E}

$$\putequation{ECT}$$
\noindent where
$$\putequation{es}$$

The momentum of the excitation is computed by means of \pl{P}
$$\putequation{PCT}$$
\noindent where
$$\putequation{ps}$$
The quantum numbers of this type of excitation are (due to the lowest
weight theorem) $S=0,\ \eta^z=-1,\ \eta =1$, and the state is lowest
weight of $SO(4)$. The complete multiplet is obtained by acting with
$\eta^\dagger$ and $(\eta^\dagger)^2$, which leaves the energy but not
the momentum invariant ($\eta^\dagger$ does not commute with the
momentum operator).

\vskip .4cm
{\sl\centerline{Charge-Singlet Excitation}}
\vskip .4cm
The charge singlet is obtained by choosing $M^{\prime}_1 = {L\over
2}-2$ and $\mp{2} =1$. The allowed range of vacancies for the integers
$J^{\prime 1}_\ga$ is $[-{L-2\over 4},{L-2\over 4}]$, so that there
are again two holes with corresponding spectral parameters
$\lh{1}$,$\lh{2}$ in the distribution. The integer $J^{\prime
2}_\ga$ can take the value $0$ only. We denote the spectral parameter
corresponding to the $k$-$\La$-string of length $2$ by $\kappa$. The
Lieb-Wu equations \pl{liebwu} for this type of excitation are
$$\putequation{BAECS}$$

In the thermodynamic limit they turn into integral equations for an
$F$-function and a density $\rho$ defined like in \pl{F} and \pl{rho}

$$\putequation{BAECS2}$$
The second equation in \pl{BAECS2} determines $\kappa$ as a function
of $\lh{j}$. The zeroth order part of this equation is fulfilled
identically, and the ${\cal O}(L^{-1})$ part leads to the requirement

$$\putequation{kappa}$$
The equation for $F_{CS}$ can then be solved by Fourier transforming
$$\putequation{FCS}$$
where $\gamma(\mu)$ is defined in \pl{fadtakh}. Energy and momentum
can be determined by using \pl{E} and \pl{P}. We find
$$\putequation{EPCS}$$
\noindent where $\ge_{cw}(\la)$ and $p_{cw}^h(\La)$ are given by
\pl{es} and \pl{ps}, and where $p^p_{cw}(\la) = \pi+p^h_{cw}(\la)$.
Thus we see that this excitation has the same energy as the charge
triplet excitation considered above, whereas the momentum differs by
$\pi$. Charge and spin quantum numbers are $\eta =0=S$.

\vskip .4cm
{\sl\centerline{Spin-Charge Scattering States}}
\vskip .4cm
We now choose $M^{\prime}_1 = {L\over 2}-1$ and ${\cal M}_e =
N_e-2\mp{}=1$. ${\cal M}_e$ is the number of ``elementary'' $k_j$'s in
a solution of \pl{pbc}-\pl{pbc2}. The number of vacancies \pl{ineq}
for the integers $J^{\prime 1}_\ga$ is ${L\over 2}$, so that there is
one hole with corresponding spectral parameter $\lh{}$ in the
distribution. The spectral parameter of the one real $k$ is denoted by
$k$. The Lieb-Wu equations read
$$\putequation{BAEes}$$
The second equation in \pl{BAEes} determines $k$ as a function of the
integer ${\tilde I}$, which has range $[-{L-1\over 2}, {L-1\over
2}]$. In the thermodynamic limit the first part of \pl{BAEes} turns
into an integral equation for an $F$-function
$$\putequation{Fes1}$$
\noindent which has the solution
$$\putequation{Fes}$$
The excitation energy of the charge-spin scattering state is given by
$$\putequation{EES}$$
\noindent where $\ge_{cw}(\la)$ is given by \pl{es} and
$$\putequation{ec}$$
The momentum is found to be
$$\putequation{PES}$$
\noindent where $p_{cw}^h(\la)$ is given by \pl{ps} and
$$\putequation{pc}$$
The quantum numbers of this excitation are $-S^z={1\over 2}=S$ and
$-\eta^z={1\over 2}=\eta$ and the state is thus the lowest weight
state of an $SO(4)$ multiplet of dimension $4$. The other states of
the multiplet are obtained by acting with $S^\dagger$, $\eta^\dagger$,
and both.
\vskip .4cm
{\sl\centerline{Spin-Singlet Excitation}}
\vskip .4cm
The spin-singlet excitation is characterized by taking $M^{\prime}_1 =
{L\over 2}-1$, ${\cal M}_e = 2$, and $M_1=1$. The total number of
electrons is $N_e=L$, so that there is no charge associated with this
excitation. The Lieb-Wu equations are
$$\putequation{BAESS}$$
Here $\kappa$ is the spectral parameter corresponding to $M_1=1$, and
$k_1, k_2$ are the momenta of the real $k$'s. There are ${L\over
2}-1$ vacancies for the half-odd integers $J^{\prime 1}_\ga$, so that
there are no holes in the distribution. The integer $J^1_\ga$ is found
to be $0$ by \pl{ineq}, which implies that $\kappa = {\sin(k_1) +
\sin(k_2)\over 2}$ by the third equation of \pl{BAESS}. The first
equation in \pl{BAESS} once again turns into an integral equation for
an $F$-function in the thermodynamic limit
$$\putequation{FSS}$$
Energy and momentum are given by
$$\putequation{EPSS}$$
where $\ge_{sw}(k)$ and $p_{sw}(k)$ are given by \pl{ec} and \pl{pc}.
The quantum numbers are $S=0=\eta$, so that the excitation is a
singlet of $SO(4)$.
\vskip .4cm
{\sl\centerline{Spin-Triplet Excitation}}
\vskip .4cm
The last type of excitation in the two-particle sector is the
spin-triplet excitation.
The quantum numbers of the lowest weight state are $\eta =0$,
$S^z=-1,\ S=1$. It is obtained by choosing
$M^{\prime}_{1} = {L\over 2}-1$ and ${\cal M}_e = 2$, which leads to a
total number of electrons $N_e={L\over 2}$. There are ${L\over 2}-1$
vacancies for the half-odd integers $J^{\prime 1}_\ga$, so that there
are no holes in the distribution. The Lieb-Wu equations are
$$\putequation{BAEST}$$
The $F$-function is found to be identical to $F_{SS}(\La)$, which
implies that energy and momentum of the spin-triplet and spin-singlet
are the same.

{\sc\section{Quasiparticle Interpretation for $U<0$}}
All of the excited states construced above have a natural
interpretation as scattering states of four elementary excitations
(``quasiparticles''), which form the fundamental representation of
$SU(2)_{spin}\times SU(2)_{eta}$, {\sl i.e.} they are grouped into
doublets of the spin and $\eta$-SU(2) respectively.
The chargeless spin-carriers (``spin-waves'') have the dispersion
$$\putequation{aepc}$$
Their quantum numbers are $\eta = 0$, $S={1\over 2}, S^z=\pm {1\over
2}$, which corresponds to the representation $({1\over 2},0)$ of
$SU(2)_{spin}\times SU(2)_{eta}$.\\
The spinless charge-carriers (``charge-waves'') (one particle, one
hole) have the dispersions\upref woynar/
$$\putequation{aeps}$$
Their quantum numbers are $S=0, \eta={1\over 2},\ \eta^z=\pm{1\over
2}$ and they form the $(0,{1\over 2})$ representation of
$SU(2)_{spin}\times SU(2)_{eta}$.\\
Like in the case of the spin-$1\over 2$ Heisenberg antiferromagnet the
quasiparticles are ``confined'', {\sl i.e.} they do not exist as
one-particle excitations. This is closely related to the fact that the
symmetry of the Hubbard hamiltonian is $SO(4)$ and all half-odd
integer representations of $SU(2)\times SU(2)$ (and thus the
fundamental representation formed by the four elementary excitations)
are not present\footnote{This is completely analogous with the
spin-$1\over 2$ Heisenberg model, which is $O(3)$-symmetric for even
chains, whereas the elementary excitations form the fundamental
representation of $SU(2)$\upref ft,ft2/.}.
The two-particle scattering states of the four quasiparticles are
easily seen to reproduce exactly the twelve excitations of section $2$.

\item{(i)} {\sl Scattering of two spin-waves}\\
The dispersion relations for the scattering states of two spin-waves
follow from \pl{aepc} to be $E=\ge_{sw}(k_1)+ \ge_{sw}(k_2)$, $P=p_{sw}(k_1)+
p_{sw}(k_2)$, where $k_1$ and $k_2$ are the rapidities of the spin-waves.
The $SO(4)$-representation is simply $({1\over 2},0)\otimes({1\over
2},0) =  (1,0)\oplus(0,0)$. By inspection we see that $(1,0)$ is
identical to the spin-triplet excitation of section $2$, and $(0,0)$ to
the spin-singlet.

\item{(ii)} {\sl Scattering of two charge-waves}\\
Scattering of two charge-waves leads to the $SO(4)$-representation
$(0,{1\over 2})\otimes(0,{1\over 2}) = (0,1)\oplus (0,0)$. The
energy of all four states follows from \pl{aeps} to be
$E=\ge_{cw}(\lh{1})+ \ge_{cw}(\lh{2})$, where $\lh{1}$ and $\lh{2}$ are the
rapidities of the charge-waves. The lowest weight state $\eta^z = -1$
of the triplet $(0,1)$ contains two holes and has momentum $P =
p_{cw}^h(\lh{1}) + p_{cw}^h(\lh{2})$. It is identical to the charge-triplet
state constructed in section $2$. The other two states of the triplet
have momenta $P_{\eta^z = 0}=p_{cw}^h(\lh{1}) + p_{cw}^p(\lh{2})$ and
$P_{\eta^z = 1}=p_{cw}^p(\lh{1}) + p_{cw}^p(\lh{2})$ respectively, which is
also in agreement with the results of section $2$. The $(0,0)$ state is
seen to be identical to the charge-singlet constructed above.

\item{(iii)} {\sl Scattering of spin-waves and charge-waves}\\
The relevant $SO(4)$-representation for this process is
$({1\over 2},0)\otimes(0,{1\over 2}) = ({1\over 2},{1\over 2})$. The
lowest weight state of the multiplet has quantum numbers $\eta =
{1\over 2} = S$, $S^z=-{1\over 2} = \eta^z$ and a dispersion
$E=\ge_{cw}(\lh{})+\ge_{sw}(k)$, $P = p_{cw}^h(\lh{}) + p_{sw}(k)$,
where $k$ and $\lh{}$ denote the rapidities of the spin-wave and
charge-wave respectively. This state is clearly identical to the
spin-charge scattering state of section $2$. The other states of the
multiplet are identical to the ones found in section $2$ as well.\\

\noindent
A similar analysis can be carried out also in the sectors with four,
six, {\sl etc.} particles. All excitations have a natural
interpretation in terms of scattering states of the four
quasiparticles. The general proof of this statement is given in the
appendix.

{\sc\section{Scattering Matrix}}

In quantum mechanical scattering theory the scattering matrix can be
extracted from the asymptotics of the wave-function of the scattering
state\upref Landau/. The boundary conditions of the quantum mechanical
problem are free. This is in contrast to the periodic boundary
conditions imposed in the Bethe Ansatz solution. In [\putref{korepin}]
it was shown how to modify quantum mechanical scattering theory to
accomodate for this fact, and a general method for extracting the
exact S-matrix from the Bethe Ansatz solution was given. In
[\putref{ft,ft2,takh,kr}] this method was applied to evaluate the exact
S-matrix for Heisenberg models. Here we will
apply this method to the case of the nested Bethe Ansatz and the
Hubbard model. The S-matrix for parity-eigenstates in one dimension is
simply a phase factor. The two-particle scattering-phase is equal to
the phase obtained by moving particle $2$ through the whole interval
(one-dimensional box) in the presence of particle $1$ {\sl minus} the
one-particle phase-shift, which is equal to the phase picked up by
moving particle $2$ through the box when $1$ is absent. The
two-particle phases can be computed from the Lieb-Wu equations
\pl{liebwu} by using our results for the two-particle excitations above.
However there are no one-particle excitations present and thus it is
impossible to evaluate the one-particle phase-shifts by means of the
Bethe Ansatz. However it is possible to evaluate the {\sl relative}
phases between two-particle excitations by means of the Bethe
Ansatz. Thus it is possible to evaluate the exact
S-matrix up to an {\sl overall} constant factor, which is the
difference between a suitably chosen reference phase, and the true
one-particle phase shift. In other words the method of
[\putref{korepin}] allows the evaluation of the logarithmic derivative
of the S-matrix. In general this leads to the subtle issue of
determining the overall phase of the S-matrix\upref melzer/, which for
models like the spin-$1\over 2$ Heisenberg antiferromagnet cannot be
resolved in the framework of the Bethe Ansatz. For the case of the
Hubbard model we are in much more fortunate position: In the limit
$U\rightarrow\infty$ the charge sector of the repulsive Hubbard model
reduces to scattering of free fermions. This can be seen by directly
evaluating the many-body wave-functions in this limit\upref woynar3/.
Thus the S-matrix (in the specific sector) must reduce the the
S-matrix for free fermions (which is $1$) in the limit
$U\rightarrow\infty$, which allows us to fix the constant.
In the $U\rightarrow\infty $ limit of the attractive Hubbard model the
charge-triplet must reduce to a scattering state of hard-core bosons
with S-matrix $-1$.\par
The complete S-matrix for the Hubbard model is $16\times 16$
dimensional and blockdiagonal. It breaks up into $4$ blocks (due to
conservation of spin and charge quantum numbers), describing
scattering of spin-waves on spin-waves, spin-waves on charge-waves,
charge-waves on spin-waves, and charge-waves on charge-waves
respectively
$$\putequation{SM}$$
\vskip .4cm
\centerline{{\sl Charge-Charge Scattering}}
\vskip .4cm
The four parity eigenstates are the charge-singlet and -triplet states.
As the scattering phase is the same for all members of the triplet we
will only consider the lowest weight state. The general form for the
S-matrix of quasiparticles in the fundamental representation of
$SU(2)$ is
$$\putequation{Sgen}$$
where $a,b,c,d$ label the colliding quasiparticles, and $\la_1$,
$\la_2$ are the corresponding spectral parameters. The phase factors
for singlet and triplet are
$$\putequation{ts}$$
We will now use the method of [\putref{korepin}] to determine $S_t$
and $S_s$.
We start by considering the charge-singlet excitation.
The phase for moving particle $2$ through the ring in presence of
particle $1$ is
$$\putequation{d12cs}$$
In order to obtain the true phase shift we now ought to subtract the
phase $\gd_{charge}(\lh{2})$ obtained by moving particle 2 through the
ring in the absence of particle 1 ({\sl i.e.} the one-particle
phase-shift). Due to confinement of the quasiparticles it is however
impossible to compute $\gd_{charge}$. Instead we will subtract a
reference phase $\gd_0(\lh{2})$, which is chosen such that the phase
$\gd_{CS}^{21}(\lh{1},\lh{2}) - \gd_0(\lh{2})$ is a function of
$\lh{2}-\lh{1}$ only. The difference $\gd_{charge} - \gd_0$ leads to
an overall phase-factor for the S-matrix \pl{SM}, which will be fixed below.
The reference phase is
$$\putequation{d0}$$
which leads to the following phase-shift for the charge-singlet
$$\putequation{dcs}$$
where $\mu_4 = {\lh{2}-\lh{1}\over 2|U|}>0$, $\theta(x) = 2\arctan(x)$,
and $\gamma(\mu)$ is defined in \pl{fadtakh}.\\

In order to determine $S_t$ we have to consider the charge triplet
excitation. The phase-shift for moving particle $2$ through the ring in
presence of particle $1$ is
$$\putequation{d12ct}$$
The extra $\pi$ is due to the fact that the $J^{\prime 1}_\ga$ are
half-odd integers for the charge-triplet but integers for the
charge-singlet and the ground state, which will be used for reference
purposes. The phase-shift is found to be
$$\putequation{dct}$$

Using \pl{dct} and \pl{dcs} we  can now determine the phase-factors
$S_s$ and $S_t$
$$\putequation{scc}$$
Thus the final result for the $4\times 4$ block of the complete
S-matrix describing scattering of Charge-waves on Charge-waves
is found to be
$$\putequation{Scc}$$
where $I$ and $\Pi$ are the $4\times 4$ identity and permutation
matrices respectively. This S-matrix is as a function of spectral
parameter (but {\sl not} of momentum) identical to the well-known
S-matrix for the spin-$1\over 2$ Heisenberg antiferromagnet\upref
ft,ft2/ and the $SU(2)_1$ Wess-Zumino-Witten-Novikov (WZWN)
model\upref wieg/ (this S-matrix also describes scattering of
particles in the $SU(2)$ Gross-Neveu model\upref natan1/, and spinons
in the Kondo model\upref natan2, natan3/).

\vskip .4cm
\centerline{{\sl Scattering of Charge on Spin and Spin on Charge}}
\vskip .4cm
The relevant excitation for these processes is the spin-charge
scattering state of section $2$. We first consider the charge-wave to
be the active scatterer.
The phase for moving the charge-wave with spectral parameter $\lh{}$
around the ring in presence of the spin-wave with spectral parameter
$k$ is
$$\putequation{d12es}$$
The reference phase is once again given by \pl{d0}, which yields the
following phase-shift for charge-spin scattering
$$\putequation{des}$$
The corresponding $4\times 4$ block of the S-matrix is
$$\putequation{Ses}$$
Now we consider the spin-wave as the active scatterer. The
total two-particle phase shift is
$$\putequation{d12se}$$
To get the phase-shift $\gd_{s\eta}$ we ought to subtract the
one-particle phase-shift for a spin-wave $\gd_{spin}(k)$, which is
unknown due to confinement. By the same method as above we can,
however, determine the phase-shift up to a constant. The reference
phase is ${\tilde\gd}_0(k) = Lk-\sum_{\gb =1}^{{L\over
2}}\theta({\sin(k) - \La_\gb\over |U|})$, which yields the following
result for $\gd_{s\eta}$
$$\putequation{dse}$$
This is {\sl a priori} correct up to a constant, which we will now
show to be zero. We know that $\gd_{\eta s}(\sin(x),x) = \gd_{s\eta
}(\sin(x),x)$, as in this case the quasiparticles are at rest relative
to each other. Inspection of \pl{dse} and \pl{des} now shows that the
constant must vanish. As a result we see that it does not matter
whether spin or charge is taken to be active. The form of the S-matrix
\pl{Ses} is the same, only the definition of $\mu$ changes.

\vskip .4cm
\centerline{{\sl Spin-Spin Scattering}}
\vskip .4cm
In the spin-spin sector the S-matrix is once again of the form
\pl{Sgen}. We first determine the phase-shift for the triplet, as the
$I_j$'s are integers both for the spin triplet and the spin-charge
scattering state we used to determine the reference phase in the
spin-sector. The phase for moving particle $2$ through the ring in
presence of particle $1$ is
$$\putequation{d12st}$$
Subtracting the one-particle phase-shift for spin-waves
${\tilde\gd}_0(k_2)$ we arrive at the following expression for the
triplet scattering phase
$$\putequation{dst}$$
The spin-singlet phase-shift is found to be
$$\putequation{dss}$$
which implies the following form for the S-matrix in the spin-spin
sector
$$\putequation{Sss}$$
Last but not least we have to fix the overall factor of the S-matrix
by considering the $U\rightarrow -\infty$ limit. In this limit the
wave-function for the charge-triplet state reduces to the one for
hard-core bosons. Therefore the charge-triplet phase-shift has to
become $\pi$ in this limit (the S-matrix for hard-core bosons is $-1$,
the one for free fermions $1$). This fixes the constant factor. We see
that our expression for the charge-triplet phase-shift \pl{dct} has
the correct asymptotic behaviour, so that our constant factor is
already the correct one. This completes our computation of the exact
S-matrix for the attractive Hubbard model. \\
The analytic structure of the S-matrix is related to the existence of
bound states of the elementary excitations. The physical strips for
spin-waves and charge-waves are defined by $|{\rm Im}(\sin(k))|\leq 2|U|$
and $|{\rm Im}(\lh{})|\leq |U|$ respectively. Using these ranges in the
expressions for the S-matrices $S_{ss}$, $S_{\eta s}$ and $S_{cc}$ we
find poles at $\sin(k_2) - \sin(k_1) = 2i |U|$ in the spin-spin sector,
at $\lh{} - \sin(k) = \pm i |U|$\footnote{The sign depends on whether
we consider the spin-wave or charge-wave to be the active scatterer.}
in the charge-spin sector, and at
$\lh{2}-\lh{1} = -2i |U|$ in the charge-charge sector.
The pole in the charge-charge sector is at the boundary of the
physical strip and corresponds to zero total energy and momentum. Thus
it does not correspond to a physical bound state. The analysis in the
spin-spin and charge-spin sectors is more difficult. For large values
of $|U|$ one can show that the poles either lie on the unphysical sheet
(and thus lead to anti-bound states, which due to non-normalizability
drop out of the physical Hilbert space), or lead to excitations with
fixed momentum (and thus cannot be interpreted as particles) and
higher energies than the scattering states. We conclude that there are
no bound states and the set of four elementary excitations is complete.

{\sc\section{Repulsive Hubbard Model}}
The analysis of the Lieb-Wu equations \pl{liebwu} for the repulsive
case is similar to the attractive case above. The analog of the
logarithmic equations \pl{pbc} - \pl{pbc2} was derived by M. Takahashi
in [\putref{taka}]. The main difference is the form of the
$k$-$\La$-strings: in the repulsive case $k^1_\ga = \pi -
\arcsin(\La^{\prime m}_\ga + i\ m U )$, $k^{2m}_\ga =\pi -
\arcsin(\La^{\prime m}_\ga - i\ m U )$, whereas all other $k^j_\ga$'s
are the same as in \pl{klstr1}(see [\putref{taka}]). Equations
\pl{pbc1b} and \pl{pbc2} are the same for the repulsive case if we
take $|U|\rightarrow U$. The analog of equation \pl{pbc} for the case
of repulsion is obtained by taking $|U|\rightarrow -U$ in \pl{pbc}.
The boundaries of integers \pl{ineq} are the same for $U>0$ and
$U<0$\footnote{This is important for the validity of the completeness
proof of [\putref{eks}] for both attraction and repulsion.}. Energy
and momentum are
$$\putequation{rE}$$
$$\putequation{rP}$$
The half-filled repulsive ground state is characterized by taking
$N_e=L$, $M_1 = {L\over 2}$, and then filling both the integers
$J^1_\ga$ and the half-odd integers $I_j$ symmetrically around zero. It is
described by a set of two coupled integral equations and was found by
Lieb and Wu in [\putref{lieb}]. The two filled Fermi seas of the
$I_j$'s and the $J^1_\ga$'s correspond to spin and charge degrees of
freedom respectively. The excitation spectrum over this ground state
has been previously studied by many authors\upref ov , woynar2 , woynar3 ,
woynar1,ksz,cy,kawa6/\footnote{For a collection of reprints see
[\putref{rv}].}, although the Lie-algebraic structure of the
excitation spectrum was not fully revealed. Like in the attractive
case there are $12$ excitations in the two-particle sector.
Unlike for the attractive case now {\sl two} $F$-functions occur as
the ground state consists of two Fermi seas. In order to explain our
notation we will discuss the charge-triplet excitation in some detail
and then quote the results for the other excitations. The Lieb-Wu
equations for the repulsive ground state are
$$\putequation{GSREP}$$
where $I_l = l-{L+1\over 2}$, $l=1\ldots L$, and  $J_\ga^1 = \ga -
{L+2\over 4}$, $\ga=1\ldots {L\over 2}$. In the thermodynamic limit
\pl{GSREP} turn into the well-know set of coupled integral
equations\upref lieb/
$$\putequation{GSREP2}$$
The ground state energy per site for the repulsive model is\upref lieb/
$$\putequation{rGSE}$$
The charge-triplet excitation over the ground state is described by
the following set of equations ($N_e=L-2,\ M_1={L-2\over 2}$)
$$\putequation{rBAECT}$$
There are two holes corresponding to spectral parameters
$\kh{1},\kh{2}$ in the distribution of the integers ${\tilde I}_j$.
The half-odd integers ${\tilde J}^1_\ga$ are distributed
symmetrically between $-{L-4\over 4}$ and ${L-4\over 4}$. Our
convention is ${\tilde I}_j - I_j = -{1\over 2}$ and ${\tilde J}^1_\ga
- J^1_\ga = -{1\over 2}$, where $I_j$ and $J^1_\ga$ are the ground state
distributions of ``integers''. Subtracting \pl{GSREP} from \pl{rBAECT}
and then taking the thermodynamic limit we can derive coupled integral
equations for the $F$-functions $F^s(\La_\ga) = {{\tilde
\La}_\ga-\La_\ga\over \La_{\ga+1}-\La_\ga}$ and $F^c(k_j) = {{\tilde
k}_j-k_j\over k_{j+1}-k_j}$
$$\putequation{rFCT}$$
These equations can once again be solved by Fourier transforming
$$\putequation{rFCT2}$$
where we have used \pl{fadtakh}.
Energy and momentum of the excitation are given by
$$\putequation{rEPCT}$$
where
$$\putequation{repc}$$
Here $\ge_h(k)$ and $p_h(k)$ are energy and momentum of a
quasiparticle carrying charge $+e$ and no spin. This quasiparticle,
which transforms in the $(0, {1\over 2})$-representation of
$SU(2)_{spin}\times SU(2)_{eta}$, is called {\sl holon}. The
charge-triplet state is a scattering state of two holons with
rapidities $\kh{1}$ and $\kh{2}$.\\

The charge-singlet is characterized by $N_e=L,\ M_1={L\over 2}-1,\
M^{\prime 1}=1$. It is a two-parameter excitation with rapidities
$\kh{1}$ and $\kh{2}$ and leads to the following set of $F$-functions
$$\putequation{rFCS}$$
Energy and momentum are found to be \upref woynar2/
$$\putequation{rPCT}$$
where $\ge_{ah}(k)\equiv \ge_h(k)$ and $p_{ah}(k) = -\pi + p_h(k)$ are
energy and momentum of the other member of the $(0, {1\over
2})$-doublet. It is connected to the holon by action of $\eta^\dagger$
and is called {\sl antiholon}. The charge-singlet is thus the
scattering state of one holon and one antiholon. \\
The spin-charge scattering state is obtained by taking $N_e=L-1,\
M_1={L\over 2}-1$ and filling the seas of integers $I_j$ and
$J^1_\ga$. There is one hole in both distributions with corresponding
spectral parameters $\lh{}$ and $\kh{}$. The $F$-functions for this
state are of the form
$$\putequation{rFes}$$
The dispersion relation is again of quasiparticle form
$$\putequation{rEetas}$$
where $\ge_s(\La)$ and $p_s(\La)$ are energy and momentum a quasiparticle
in the $({1\over 2}, 0)$-representation of $SU(2)_{spin}\times
SU(2)_{eta}$, which is called {\sl spinon}
$$\putequation{reps}$$
Spin-triplet and singlet states are scattering states of two spinons.
The lowest weight state of the triplet is obtained from the Bethe
Ansatz by taking $N_e=L,\ M_1 = {L\over 2}-1$. There are two holes
with spectral parameters $\lh{1}\ ,\lh{2}$ in the distribution of the
$\La_\ga$. The $F$-functions are found to be
$$\putequation{rFST}$$
The spin-singlet state is characterized by $N_e=L,\ M_1 = {L\over
2}-2,\ M_2=1$. Again there are two holes, and the $F$-functions are
$$\putequation{rFSS}$$
Energy for both triplet and singlet states are of quasiparticle form
$$\putequation{rEPspin}$$
where $\ge_s(\la)$ and $p_s(\la)$ are given by \pl{reps}.
The two-particle S-matrix for the repulsive case can be computed by
the same method as for the attractive case. The phase-shifts are found
to be
$$\putequation{rphase1}$$
$$\putequation{rphase2}$$
$$\putequation{rphase3}$$
Here we have fixed the over all constant by considering the
charge-triplet in the $U\rightarrow\infty$ limit, in which the wave
functions reduce to the ones for free fermions. Thus the S-matrix must
reduce to $1$ in this limit. The complete S-matrix is then again of
form \pl{SM}, where now
$$\putequation{rSss}$$
$$\putequation{rSse}$$
$$\putequation{rSes}$$
$$\putequation{rScc}$$
The analytic structure of the repulsive S-matrix can easily be
inferred from the structure of the attractive one. There are again no
{\sl physical} poles of the S-matrix in the complex plane.

{\sc\section{Repulsion versus Attraction}}
Attractive and repulsive quasiparticle spectra are related by an
interchange of spin and charge degrees of freedom. To see this we
first make the substitution $k\rightarrow \pi - k$ in the attractive
case. This renaming of spectral parameters has of course no physical
consequences. The spinon energy \pl{reps} for $U>0$ is the same as the
energy \pl{aeps} of the charge waves in the attractive case. The same
holds for the holon energy \pl{repc} and the spin-wave energy
\pl{aepc}. On the level of the quasiparticle momenta these
equivalences hold only up to constants. The spinon momentum is related
to the momentum of the charge waves for $U<0$ like $p_s(\la) =
p^h_{cw}(\la)+ \p2=p^p_{cw}(\la)- \p2$. The spin-wave momentum
$p_{sw}(k)$ is almost the same as the holon/antiholon momenta for
$U>0$ $p_{cw}(k)=p_{h}(k)+ \p2=p_{ah}(k)- \p2$.
These differences in the quasiparticle momenta are due to the fact
that the transformation $c^\dagger_{j,\up}\leftrightarrow
(-1)^jc_{j,\up}$ that interchanges spin and charge degrees of freedom
does not commute with the momentum operator but changes the total
momentum by $\pi$.
This can also be easily seen on the level of the two $SU(2)$ algebras:
the spin-$SU(2)$, which commutes with total momentum, is transformed
into the $\eta$-$SU(2)$, which changes momentum by $\pi$.\\
The S-matrices for attraction and repulsion after interchange of spin
and charge labels are seen to be identical as functions of the
uniformizing parameter $\mu$.

{\sc\section{Discussion}}
In this paper we derived the quasiparticle interpretation for the
one-dimensional Hubbard model and evaluated the exact S-matrix
describing the scattering of these quasiparticles.
The classification of elementary excitations into spinons (which carry
spin but no charge) and holons/antiholons (which carry charge but no
spin) clearly reflects spin-and charge separation in the
one-dimensional Hubbard model\footnote{The analogous statement for
spin-waves and charge-waves in the attractive case is of course also
true.}. The S-matrix is a $SU(2)\times SU(2)$-invariant solution of
the Yang-Baxter equation, which
establishes the one-dimensional Hubbard model as a factorizable
$SO(4)$-scattering theory. Our S-matrix is very similar to the
S-matrix for the $SU(2)\times SU(2)$ principal chiral model with
Wess-Zumino term of level $1$\upref zamo/. The charge-charge and
charge-spin sectors are also identical to the S-matrix for
electron-electron and electron-kink scattering in the ``exactly
screened case'' of the Kondo problem\upref fend/. Apart from its
relevance in the field-theory limit the S-matrix is an important
tool in the study of transport properties\upref Carmelo, Carmelo2/,
and plays a crucial role in the general theory of integrable models,
where S-matrix and R-matrix are often proportional. In
[\putref{shastry}] a R-matrix for an embedding of the Hubbard
hamiltonian in a commuting family of transfer matrices was derived.
This R-matrix, which belongs to a {\sl fundamental model} (where
L-operator and R-matrix are essentially identical\upref vladb/), is
different from our S-matrix. It is quite natural to search for a
different embedding of the Hubbard hamiltonian, using an R-matrix
proportional to our S-matrix. The L-operator for such a construction
would have to be different from the R-matrix.

\vskip .3cm
\centerline{\sc Acknowledgements:}
\vskip .2cm
It is a pleasure to thank Prof. C.N. Yang and Dr. E. Melzer for
stimulating discussions. This work was supported in part by the
National Science Foundation under grant 9309888 and by NATO
9.15.02 RG 901098 Special Panel on Chaos Order and Patterns
`Functional Integral Methods in Statistical Mechanics and
Correlations'.
\sectionnumstyle{Roman}
\sectionnum=0
{\sc\section{Appendix: $2N$-Quasiparticle Sector}}
In this appendix we use the Thermodynamic Bethe Ansatz (TBA)\upref
yaya/\footnote{For a review see [\putref{nepo}].} to prove
the validity of the quasiparticle interpretation (obtained by
analyzing the two-particle sector) in the $2N$-quasiparticle sector.
We will discuss the $U<0$ case explicitly, the repulsive case can be
treated in an analogous manner. The
TBA-part of the analysis in the attractive case discussed here is very
similar to the one for repulsion, which is presented in great detail in
[\putref{taka}]. We refer to that paper for explanations and notations
and merely give the results here. Starting point for the TBA are the
Bethe equations \pl{pbc} - \pl{pbc2}. In the thermodynamic limit these
equations turn into coupled integral equations for the
particle-densities $\rho$, $\sigma_n$, $\sigma^\prime_n$ and hole-densities
$\bar\rho$, ${\bar\sigma}_n$, ${\bar\sigma}^\prime_n$
($n=1,\ldots\infty$) describing the distribution of elementary $k$'s,
$\La$-strings of length $n$, and $k-\La$-strings of length $n$
respectively
$$\putequation{holes}$$
where
$$\eqalign{ A_{nm}*f\bigg|_\la &= \delta_{nm} f(\la) +{1\over 2\pi}
{d\over d\la} \int_{-\infty}^{\infty} d\la^\prime
\theta_{n,m}({\la-\la^\prime\over |U|})\ f(\la^\prime)\quad .\cr}$$
More precisely, \pl{holes} determine the hole-densities in terms of
the particle densities. The particle densities (at finite temperature)
are determined by minimizing the thermodynamic potential $\Omega =
E-TS-\mu N$ ($\mu$ is the chemical potential\footnote{We will set
$\mu=2U$ as we are interested in the half-filled case.}, $T$ the
temperature, $S$ the entropy, and $E$ the energy following from
\pl{E}). This leads to the following equations for the ratios of
densities $\zeta(k) =
{{\bar\rho}(k)\over\rho(k)}$, $\eta_n(\La) = {{\bar\gs_n}(\La)
\over\gs_n(\La)}$, and $\eta^\prime_n(\La) =
{{\bar\gs^\prime_n}(\La)\over \gs^\prime_n(\La)}$
$$\putequation{TBA}$$
The ratios of densities are related to the dressed energies
$\kappa(k)$ (for elementary $k$'s), $\ge_n(\La)$ (for $\La$-strings of
length $n$) and $\ge^\prime_n(\La)$ (for $k-\La$-strings of length
$n$) {\sl via} $\zeta(k) = \exp({\kappa(k)\over T})$, $\eta_n(\La) =
\exp({\ge_n(\La)\over T})$, and $\eta^\prime_n(\La) =
\exp({\ge^\prime_n(\La)\over T})$. In the zero-temperature limit all
dressed energies except $\ge_1^\prime(\La)$ are greater or equal to
zero, whereas $\ge_1^\prime(\La)\leq 0$ on the whole real line. This
implies that the attractive ground state is obtained by filling all
vacancies for $k-\La$-strings of length $1$, which is how we
constructed it in section $2$. In the $T=0$ limit it is also possible
to completely solve the system \pl{TBA} for the dressed energies.
We find
$$\putequation{dresseden}$$
This is in perfect agreement with our analysis of section $2$: making
a hole at spectral parameter $\lh{}$ in the distribution of
$\gs_1^\prime(\La)$ costs energy $-\ge_1^\prime(\lh{})$. This
obviously corresponds to a charge-wave, and we identify $\ge_1^\prime
= -\ge_{cw}$. Introducing one elementary $k$ increases the energy by
$\kappa(k)$, which is seen to be identical to the spin-wave energy
$\ge_{sw}(k)$. The additional information we get out of \pl{dresseden}
is that {\sl only} these two dressed energies are nonvanishing. Thus
the energy of {\sl any} excited state is the sum over $-\ge_1^\prime$'s
and $\kappa$'s. A similar analysis can be carried out for the total
momentum of excitations at $T=0$. Using the expression
$$ P = {2\pi\over L}\left(\sum_{j=1}^L I_j -
\sum_{n=1}^\infty\sum_{\ga=1}^{M_n}J^n_\ga +
\sum_{n=1}^\infty\sum_{\ga=1}^{M^\prime_n}J^{\prime n}_\ga \right)
+\pi\sum_{n=1}^\infty M^\prime_{2n}$$
for the total momentum it can be shown that only $k-\La$-strings of
length $1$ and elementary $k$'s contribute dynamically
($k-\La$-strings of lengths greater than $1$ contribute $\pi (n-1)$).
Thus we conclude that both energy and momentum of any excitation
breaks up into sums over the quasiparticle energies and momenta. What
remains to be shown is now that the quasiparticle interpretation (QPI)
predicts the correct number of $SO(4)$-representations with spin $S$ and
$\eta$-spin $\eta$ in the sector with $2N$ quasiparticles (recall that
there always are an even number).
The prediction of the QPI for the number of lowest-weight scattering
states of ${\cal M}_e$ spin-waves with spin $S$ (note that there is a
constraint that ${{\cal M}_e\over 2}-S$ is always an integer) is
simply
$$\chi_1 = {{\cal M}_e\choose {{\cal M}_e\over 2}-S} -{
{\cal M}_e \choose {{\cal M}_e\over 2}-S-1}.$$
This number has to be multiplied by the number of lowest-weight
scattering states of $2N-{\cal M}_e$ charge-waves, which is
$$\chi_2 = {2N-{\cal M}_e \choose N-{{\cal M}_e\over 2}-\eta} -
{2N-{\cal M}_e \choose N-{{\cal M}_e\over 2}-\eta-1} .$$
Are these numbers reproduced by the Bethe Ansatz ? The answer is of
course ``Yes''. Excitations over the ground state are construced based
on the allowed ranges of integers \pl{ineq}. By analyzing these
equations we can determine the exact number of Bethe Ansatz
excitations ({\sl i.e.} lowest weight states of $SO(4)$ multiplets).
We fix the number of holes in the $\La^{\prime 1}$ distribution to be
$2N-{\cal M}_e$ and the number of elementary $k$'s to be ${\cal M}_e$.
The dispersion of such states is by the above arguments identical to
the dispersion of a scattering state of ${\cal M}_e$ spin-waves and
$2N-{\cal M}_e$ charge-waves. If we further fix the values $S$ of spin
and $\eta$ of $\eta$-spin, the total number of Bethe states with these
characteristica is given by
$$\eqalign{&
\left\lbrack\sum_{{M_1,M_2,\ldots\atop\scr \sum_{m=1}^\infty mM_m =
{{\cal M}_e\over 2}-S}}\prod_{n=1}^\infty{{\cal M}_e -
\sum_{m=1}^\infty t_{mn}M_m\choose M_n}\right\rbrack
\times\cr
&\times \left\lbrack\sum_{{M^\prime_2,M^\prime_3,\ldots\atop\scr
\sum_{m=2}^\infty (m-1)M^\prime_m ={N-{\cal M}_e\over
2}-\eta}}\prod_{n=2}^\infty{2N-{\cal M}_e -
\sum_{m=2}^\infty (t_{mn}-2)M^\prime_m\choose
M^\prime_n}\right\rbrack .\cr} $$
These sums can simply be looked up in [\putref{eks}] or
[\putref{takaxxx}] and we find that the two factors exactly coincide
with $\chi_1$ and $\chi_2$. This establishes the validity of the QPI
in the $2N$-quasiparticle sector.
\begin{putreferences}
\centerline{{\sc References}}
\smallfonts
\vskip .5cm

\reference{pwa}{P.W. Anderson,\ \sci{235}{1987}{1196}.}
\reference{pwa2}{G. Baskaran, Z. Zou, P.W. Anderson,\
\SSC{63}{1987}{973}.}
\reference{lieb}{E.H. Lieb, F.Y. Wu,\ \PRL{20}{1968}{1445}.}
\reference{hl}{O.J. Heilmann, E.H. Lieb,\ \ANYAS{172}{1971}{583}.}
\reference{yang}{C.N. Yang,\ \PRL{19}{1967}{1312}.}
\reference{yang1}{C.N. Yang,\ \PRL{63}{1989}{2144}.}
\reference{yang2}{C.N. Yang and S. Zhang,\ \MPLB{4}{1990}{759}.}
\reference{ft}{L.D. Faddeev, L. Takhtajan, \JSM{24}{1984}{241}.}
\reference{ft2}{L.D. Faddeev, L. Takhtajan,\ \PLA{85}{1981}{375}.}
\reference{takh}{L. Takhtajan,\ \PLA{87}{1982}{479}.}
\reference{woynar}{F. Woynarovich,\ \JPC{16}{1983}{6593}.}
\reference{woynar1}{F. Woynarovich,\ \JPC{16}{1983}{5293}.}
\reference{woynar2}{F. Woynarovich,\ \JPC{15}{1982}{85}.}
\reference{woynar3}{F. Woynarovich,\ \JPC{15}{1982}{97}.}
\reference{vladb}{V.E. Korepin, G. Izergin and N.M. Bogoliubov,\ {\sl
Quantum Inverse Scattering Method, Correlation Functions and Algebraic
Bethe Ansatz}, Cambridge University Press, 1993}
\reference{korepin}{V.E. Korepin,\ \TMP{76}{1980}{165}.}
\reference{eks}{F.H.L. E\char'31ler, V.E. Korepin, K. Schoutens,\
\NPB{372}{1992}{559}, \NPB{384}{1992}{431}, \PRL{67}{1991}{3848}.}
\reference{Landau}{L.D. Landau, E.M. Lifshitz, ``{\sl Quantum
Mechanics}'', Pergamon Press 1975.}
\reference{ov}{A.A. Ovchinnikov,\ \JETP{30}{1970}{1160}.}
\reference{taka}{M. Takahashi,\ \PTP{47}{1972}{69}.}
\reference{sm}{F.H.L. E\char'31ler, V.E. Korepin,\ {\sl SO(4)
Invariant Scattering Matrix of the Hubbard Model, preprint}}
\reference{Carmelo}{J. Carmelo, A.A. Ovchinnikov,\
\JPCM{3}{1991}{757}.}
\reference{Carmelo2}{J. Carmelo, P. Horsch, A.A. Ovchinnikov,\
\PRB{46}{1992}{14728}.}
\reference{shastry}{B.S. Shastry,\ \JSP{50}{1988}{57}.}
\reference{rv}{{\sl Exactly Solvable Models of Strongly Correlated
Electrons}, eds V.E. Korepin, F.H.L. E\char'31ler,\ World Scientific,
{\sl to appear}}
\reference{melzer}{T.R. Klassen, E. Melzer, {\sl preprint ITP-SB-92-36}.}
\reference{ksz}{A. Kl\"umper, A. Schadschneider, J. Zittartz,\
\ZPB{78}{1990}{99}.}
\reference{cy}{T.C. Choy, W. Young,\ \JPC{15}{1982}{521}.}
\reference{kawa6}{N. Kawakami, A. Okiji,\ \PRB{40}{1989}{7066}.}
\reference{suth}{B.Sutherland,\ ,\ {\sl
Exactly Solvable Problems in Condensed Matter and Relativistic field
Theory},\ B.S. Shastry, S.S. Jha, V. Singh (eds.)\ Lecture Notes in
Physics, v.242, Berlin: Springer Verlag, (1985), p.1 .}
\reference{bik2}{V.E. Korepin, G. Izergin and N.M. Bogoliubov,\ {\sl
Exactly Solvable Problems in Condensed Matter and Relativistic field
Theory},\ B.S. Shastry, S.S. Jha, V. Singh (eds.)\ Lecture Notes in
Physics, v.242, Berlin: Springer Verlag, (1985), p.220.}
\reference{wieg}{P.B. Wiegmann,\ \PLB{141}{1984}{217}.}
\reference{yaya}{C.N. Yang, C.P. Yang,\ \JMP{10}{1969}{1115}.}
\reference{nepo}{L. Mezincescu, R.I. Nepomechie, {\sl preprint
UMTG-170}.}
\reference{takaxxx}{M. Takahashi,\ \PTP{46}{1971}{401}.}
\reference{taka2}{M. Takahashi,\ \PTP{45}{1971}{756}.}
\reference{per}{M. Pernici,\ \EPL{12}{1990}{75}.}
\reference{aff}{I. Affleck in talk given at the Nato Advanced
Study Institute on {\sl Physics, Geometry and Topology}, Banff, August
1989.}
\reference{Carmelo}{J. Carmelo, A.A. Ovchinnikov,\ \JPCM{3}{1991}{757}.}
\reference{shastry}{B.S. Shastry,\ \JSP{50}{1988}{57}.}
\reference{rv}{{\sl Exactly Solvable Models of Strongly Correlated
Electrons}, eds V.E. Korepin, F.H.L. E\char'31ler,\ World Scientific,
{\sl Spring 1994}}
\reference{fend}{P. Fendley,\ {\sl Kinks in the Kondo-problem},\ {\sl
preprint} BUHEP-93-10.}
\reference{zamo}{A.B. Zamolodchikov, Al.B. Zamolodchikov,\
\NPB{379}{1992}{602}.}
\reference{kr}{A.N. Kirillov, N.Yu. Reshetikhin,\ \JPA{20}{1987}{1565}.}
\reference{natan1}{N. Andrei, J.H. Lowenstein,\ \PLB{91}{1980}{401}.}
\reference{natan2}{N. Andrei, summer course on {\sl Low-dimensional
Quantum Field Theories for Condensed Matter Physicists}, Trieste 1992,
unpublished.\\
These lecture notes, which were brought to our attention
after completion of our paper, also discuss the computation of
phase-shifts in the Hubbard model.}
\reference{natan3}{N. Andrei, {\sl private communication}.}
\end{putreferences}
\bye